\documentstyle{mn}

\newif\ifAMStwofonts

\ifoldfss
  \ifCUPmtlplainloaded \else
    \NewTextAlphabet{textbfit} {cmbxti10} {}
    \NewTextAlphabet{textbfss} {cmssbx10} {}
    \NewMathAlphabet{mathbfit} {cmbxti10} {} 
    \NewMathAlphabet{mathbfss} {cmssbx10} {} 
  \fi
  \ifAMStwofonts
    \ifCUPmtlplainloaded \else
      \NewSymbolFont{upmath} {eurm10}
      \NewSymbolFont{AMSa} {msam10}
      \NewMathSymbol{\upi}     {0}{upmath}{19}
      \NewMathSymbol{\umu}     {0}{upmath}{16}
      \NewMathSymbol{\upartial}{0}{upmath}{40}
      \NewMathSymbol{\leqslant}{3}{AMSa}{36}
      \NewMathSymbol{\geqslant}{3}{AMSa}{3E}

      \let\geq=\geqslant 
    \fi
  \fi
\fi 

\ifnfssone
  \newmathalphabet{\mathit}
  \addtoversion{normal}{\mathit}{cmr}{m}{it}
  \addtoversion{bold}{\mathit}{cmr}{bx}{it}
  \newmathalphabet{\mathbfit} 
  \addtoversion{normal}{\mathbfit}{cmr}{bx}{it}
  \addtoversion{bold}{\mathbfit}{cmr}{bx}{it}
  \newmathalphabet{\mathbfss} 
  \addtoversion{normal}{\mathbfss}{cmss}{bx}{n}
  \addtoversion{bold}{\mathbfss}{cmss}{bx}{n}
  \ifAMStwofonts
    \ifCUPmtlplainloaded \else
      \UseAMStwoboldmath
      \makeatletter
      \new@mathgroup\upmath@group
      \define@mathgroup\mv@normal\upmath@group{eur}{m}{n}
      \define@mathgroup\mv@bold\upmath@group{eur}{b}{n}
      \edef\UPM{\hexnumber\upmath@group}
      \new@mathgroup\amsa@group
      \define@mathgroup\mv@normal\amsa@group{msa}{m}{n}
      \define@mathgroup\mv@bold\amsa@group{msa}{m}{n}
      \edef\AMSa{\hexnumber\amsa@group}
      \makeatother
      \mathchardef\upi=''0\UPM19
      \mathchardef\umu=''0\UPM16
      \mathchardef\upartial=''0\UPM40
      \mathchardef\leqslant=''3\AMSa36
      \mathchardef\geqslant=''3\AMSa3E

      \let\geq=\geqslant 
    \fi
  \fi
\fi 

\ifnfsstwo
  \DeclareMathAlphabet{\mathbfit}{OT1}{cmr}{bx}{it}
  \SetMathAlphabet\mathbfit{bold}{OT1}{cmr}{bx}{it}
  \DeclareMathAlphabet{\mathbfss}{OT1}{cmss}{bx}{n}
  \SetMathAlphabet\mathbfss{bold}{OT1}{cmss}{bx}{n}
  \ifAMStwofonts
    \ifCUPmtlplainloaded \else
      \DeclareSymbolFont{UPM}{U}{eur}{m}{n}
      \SetSymbolFont{UPM}{bold}{U}{eur}{b}{n}
      \DeclareSymbolFont{AMSa}{U}{msa}{m}{n}
      \DeclareMathSymbol{\upi}{0}{UPM}{``19}
      \DeclareMathSymbol{\umu}{0}{UPM}{``16}
      \DeclareMathSymbol{\upartial}{0}{UPM}{``40}
      \DeclareMathSymbol{\leqslant}{3}{AMSa}{``36}
      \DeclareMathSymbol{\geqslant}{3}{AMSa}{``3E}

      \let\geq=\geqslant 
    \fi
  \fi
\fi 

\ifCUPmtlplainloaded \else
  \ifAMStwofonts \else 
    \def\upi{\pi}
    \def\umu{\mu}
    \def\upartial{\partial}
  \fi
\fi

\title[Mass-to-light ratios]
      {Mass-to-light ratios from the fundamental plane of spiral galaxy disks}
\author[Alister W. Graham]
       {Alister W.~Graham\thanks{Present Address: Department of Astronomy, 
       University of Florida, Gainesville, FL, USA}\\
        Instituto de Astrof\'{i}sica de Canarias, La Laguna, E-38200, Tenerife, 
        Spain}
\date{Accepted 2002 January 23}

\pagerange{\pageref{firstpage}--\pageref{lastpage}}
\pubyear{2001}

\begin{document}
\input{psfig}

\label{firstpage}

\maketitle

\begin{abstract}

The best-fitting 
2-dimensional plane within the 3-dimensional space of spiral galaxy disk 
observables (rotational velocity $v_{\rm rot}$, central disk surface 
brightness $\mu_0 = -2.5\log I_0$, and disk scale-length $h$) has 
been constructed. 
Applying a three-dimensional bisector method of regression analysis
to a sample of $\sim$100 spiral galaxy disks that span more than four 
mag arcsec$^{-2}$ in central disk surface brightness yields 
$v_{\rm rot}\propto I_0^{0.50\pm0.05}h^{0.77\pm0.07}$ ($B$-band) and
$v_{\rm rot}\propto I_0^{0.43\pm0.04}h^{0.69\pm0.07}$ ($R$-band). 
%
%
%
Contrary to popular belief, these results suggest that in the $B$-band, 
the dynamical mass-to-light ratio (within 4 disk scale-lengths) is largely 
independent of surface brightness, varying as 
$I_0^{0.00\pm0.10}h^{0.54\pm0.14}$.  
Consistent results were obtained when the expanse of the 
analysis was truncated by excluding the low surface brightness galaxies. 
{\it Previous claims that $M/L_B$ varies with $I_0^{-1/2}$ are 
shown to be misleading and/or due to galaxy selection effects. 
Not all low-surface-brightness disk galaxies are dark matter 
dominated.} 
The situation is however different in the near-infrared where 
$L_{K'}\propto v^4$ and $M/L_{K'}$ is shown to vary as $I_0^{-1/2}$. 
{\it Theoretical studies of spiral galaxy disks should not assume a 
constant $M/L$ ratio within any given passband.} 

The $B$-band dynamical mass-to-light ratio (within 4 disk scale-lengths) 
has no obvious correlation with ($B-R$) disk colour, while in the 
$K'$-band it varies as -1.25$\pm$0.28($B-R$). 
Combining the present observational data with recent galaxy model 
predictions implies that the logarithm of the stellar-to-dynamical 
mass ratio is not a constant value, but increases as disks become 
redder, varying as 1.70$\pm0.28$($B-R$).

\end{abstract}


\section{Introduction}

Given the large observed range in the disk parameters of spiral galaxies, 
the tightness (and slope) of the Tully \& Fisher (1977, hereafter TF) 
relation between absolute magnitude and the logarithm of the rotational 
velocity 
is a fundamental clue as to how these systems were constructed
(Dalcanton, Spergel \& Summers 1997; Mo, Mao \& White 1998; 
Mo \& Mao 2000; van den Bosch 2000; Koda, Sofue \& Wada 2000).  
Not only does this relationship hold for the High Surface Brightness
(HSB) spiral galaxies that typify the Hubble sequence, but also for 
the disk-dominated Low Surface Brightness (LSB) spiral galaxies 
(Zwaan et al.\ 1995).  
The universality of the TF relation for both (HSB) and LSB galaxies 
immediately tells us that for a given rotational velocity one has 
a constant galaxy luminosity, albeit within the scatter of the TF 
relation.  This had been used to infer that, at any given constant 
luminosity, the mass-to-light ($M/L$) ratio must vary 
linearly with increasing disk scale-length $h$, or equivalently 
as $I_0^{-1/2}$, where $I_0$  is the central disk surface brightness
in intensity units (Zwaan et al.\ 1995).  Despite popular misconception,
how the $M/L$ ratio may 
vary between galaxies having different luminosities was not addressed 
and is the subject of the present analysis.  This paper derives the mean 
dependency of the $M/L$ ratio on the scale-length and (central) surface 
brightness of a galaxy's disk. 

The issue above is tackled by replacing the magnitude term in the 
TF diagram with the central disk intensity and 
scale-length terms (recall, $L_{\rm disk}=2\pi I_0h^2$).  
This gives three fundamental variables:  the rotational velocity 
of the disk ($v_{\rm rot}$), the central disk surface brightness 
($\mu_0=-2.5\log I_0$ mag arcsec$^{-2}$), and the disk scale-length ($h$).  
Just as the Faber-Jackson (1976) relation between 
absolute magnitude and velocity dispersion for Elliptical galaxies 
was supplemented through the construction of the `Fundamental Plane' 
(Djorgovski \& Davis 1987; Dressler et al.\ 1987; Faber et al.\ 1987), 
this study will explore the question of whether or not the use of three 
variables, rather than two are required for the treatment of spiral 
galaxy disks.  

In recent years, numerous authors have addressed this question using 
different galaxy samples and different methods of data analysis 
(e.g., Willick et al.\ 1997).   
It should be noted that this question is approached here 
by treating all three variables equally (Feigelson \& Babu 1992). 
It must therefore be understood that what is meant in the following text 
by `departure from the TF relation' is a departure of the optimal 
`fundamental plane' of spiral galaxy disks from the `TF plane'
when all three disk parameters are treated equally. 
Relations constructed for the purpose of galaxy distance determinations 
(see, for e.g., Triay, Lachieze-Rey \& Rauzy 1994), 
where, typically, the residuals of only one variable about some 
relation are minimised are different; the results of such studies 
cannot be readily compared with the current investigation. 
Consequently, a number of such published data sets are reanalysed 
here in a consistent fashion. 
The tilt to the optimal `fundamental plane' of spiral galaxy disks 
is subsequently used to infer the mass-to-light ratio as a function 
of disk scale-length and surface brightness. 

Section 2 introduces the field galaxy samples that have been 
used, and the techniques employed to obtain their disk parameters. 
Section 3 describes the two-dimensional plane which is used to 
represent the location of galaxy disks in the three-dimensional 
space of observables, and describes how to relate this plane to 
the $M/L$ ratio. 
The results and analysis of fitting such a plane 
to the data sample in hand is presented in section 4.   
The same method of analysis is then
applied to various literature data samples, and extended to longer 
wavelengths through the study of a homogeneous sample of Cluster 
spiral galaxies for which $B, R, I$, and $K'$ data is available.
Lastly, a brief inspection of the $M/L$ 
ratio dependency on disk colour is performed. 
The main results from this study are given in Section 6.

\section{Photometric and spectroscopic data} 

Three photometric samples have been compiled, containing 
galaxy profiles in both $B$- and $R$-bands.  All these galaxies 
were then structurally modelled using the same algorithm. 
In order to cover a large range of disk parameters, the 
well-defined sample of morphologically undisturbed HSB galaxies from 
de Jong \& van der Kruit (1994) and de Jong (1996) 
was used together with 
two large samples of LSB galaxies for which CCD 
data are available.  The first of which is the sample of 
21 disk-dominated LSB spiral galaxies from de Blok et al.\ (1995). 
The second is the sample of 21 bulge-dominated LSB spiral galaxies 
from Beijersbergen, de Blok \& van der Hulst (1999). 

The selection criteria are discussed at length in each paper, 
and they have been previously summarised in Graham \& de Blok (2001).  
Briefly, 
de Jong \& van der Kruit (1994) constructed 
a statistically complete, diameter- and volume-limited sample of 86 
HSB disk galaxies having a red minor-to-major ratio greater than 0.625 
(i.e.\ inclinations less than $\sim$50$^{\rm o}$), and a red 
major-axis $\geq$2.0\arcmin.  
Larger and brighter disk galaxies are of course over-represented in 
such a sample as they can be detected at greater distances; 
equivalently, small faint disks will be under-represented. 
Following McGaugh \& Blok's (1998) study of LSB 
galaxies and the TF relation, the current objective is simply to 
expand the baseline of observable parameters which are fed into 
such an analysis.  
This HSB sample is therefore supplemented here with two LSB galaxy 
samples; 
however, as it turns out, the main conclusions from this study 
are supported by the analysis of the HSB data alone. 

The first LSB sample is from de Blok et al.\ (1995), consisting 
of late-type LSB galaxies from van der Hulst et al.\ (1993) 
and the lists by Schombert \& Bothun (1988) and Schombert et al (1992).
De Blok et al.\ randomly selected a sub-set of galaxies with 
inclinations less than 60$^{\rm o}$, central surface 
brightnesses fainter than 23 B-mag arcsec$^{-2}$, and having 
single-dish H$_{\rm I}$ observations available. 
The LSB sample from Beijersbergen et al.\ (1999) is a random selection 
of `bulge-dominated' LSB disk galaxies 
imaged in the ESO-LV catalog (Lauberts 
\& Valentijn 1989).  Galaxies were selected if they had: inclinations 
less than 50$^{\rm o}$, diameters of the 26 B-mag arcsec$^{-2}$ 
isophote greater than 1$\arcmin$ and smaller than 3$\arcmin$ 
(chip size limitation), and a surface brightness of the disk at the 
half-light radii fainter than 23.8 B-mag arcsec$^{-2}$. 
The combined sample of HSB and LSB galaxies is not complete 
in any volume-limited sense, but does span more than 4 mag 
arcsec$^{-2}$ in central disk surface brightness.  

Every profile has been modelled by simultaneously fitting a 
seeing-convolved S\'ersic (1968) $r^{1/n}$ bulge and a 
seeing-convolved exponential disk.  Uncertainty 
in the relative sky-background level can be a significant source of 
error when determining the disk parameters; consequently, the 
parameters for the LSB galaxies ($\sim$ 1/3 of the final sample) 
are less secure than those for the HSB galaxies.  The LSB galaxies 
have typical errors of $\sim$20 per cent in their disk scale-length 
determinations, while the HSB galaxies generally have 
errors of less than 10 per cent (and typically 7 per cent).  The profile 
derived central surface brightness measurements are accurate to typically 
0.1 mag arcsec$^{-2}$ for all galaxy types. 
They have also all been homogeneously corrected for: Galactic extinction 
(Schlegel, Finkbeiner \& Davis 1998), $K$-correction (using the tables 
of Poggianti 1997), $(1$$+$$z)^4$ redshift dimming, and inclination
corrected following the prescription Tully \& Verheijen (1997) 
derived for the Ursa Major galaxy Cluster.  

The inclination of each galaxy (determined from the $R$-band images) 
have been taken from their respective papers, with the exception that 
the inclinations given in de Blok, McGaugh \& van der Hulst 
(1996) were used for the disk-dominated LSB galaxy sample. 
The galaxies have low to intermediate inclinations ($i$$<$$60^\circ$). 
On the one hand, this is good because it means that the 
required surface brightness inclination corrections (to a face-on 
orientation) are small.  Failing to account for such 
internal extinction can result in substantial errors -- 
these can be greater than 2 mag for highly inclined disks. 
On the other hand, it means that the inclination corrections (to an 
edge-on orientation) for the rotational velocity terms will be large. 

The 21 cm velocity data has come from a number of sources.
The linewidth at 20 per cent of the peak HI intensity, denoted by
$W_{20}$ and usually measured to an accuracy of a few percent, 
has been used in all cases.  Following Zwaan et al.\ (1995), 
these values were corrected for both inclination and random 
motion effects, as detailed in Tully \& Fouqu\'e (1985), to 
give the rotational velocity $v_{\rm rot}$. 
Dynamical data for the disk-dominated LSB galaxy sample has come
from de Blok et al.\ (1996) and van der Hulst et al.\ (1993), 
and is thoroughly described in these papers. 
The data for the bulge-dominated LSB galaxies has come from
de Blok (2001, priv. comm.) and was reduced in the same way as the 
disk-dominated 
sample of LSB galaxies.  Measurements of $W_{20}$ for the HSB 
galaxy sample have come from either the HI catalog of Huchtmeier 
(2000) or the database of LEDA.  Uncertainties in the true 
inclination of each disk (due to possible intrinsic disk elongation) 
is likely to be the dominant source of error in the rotational 
velocity measurements and will contribute to scatter in the 
following diagrams.  
Adding an error estimate of 10$^\circ$ to every galaxies 
photometrically derived inclination angle resulted in a mean 
change of 0.09 dex in the derived rotational velocities.  
If more accurate inclinations derived from kinematical data
(see, e.g., Franx \& de Zeeuw 1992; Andersen et al.\ 2001) 
were available, one could expect the scatter in the following 
relations to be reduced. 
Similarly, rotation curve asymmetries have not been dealt with
(Richter \& Sancisi 1994; Haynes et al.\ 1994; Matthews \& Gallagher 2002).


Considering galaxies with both photometric and kinematic data 
left a sample of 65 HSB and 30 LSB galaxies in the $B$-band, 
and 64 HSB and 31 LSB galaxies in the $R$-band (see Table 1). 
The galaxy F564-v3, the closest galaxy in the sample of disk-dominated 
LSB galaxies -- at a distance of only 6 Mpc -- is the only galaxy 
excluded at this point due to the uncertainty on its true 
distance\footnote{Inclusion of F564-v3 in the subsequent analysis 
reveals that it lies far from the relation defined by the other 94 
galaxies in the sample; suggesting its distance is not well indicated 
by its redshift.}.   

\begin{table*}
\centering
\begin{minipage}{140mm}
\caption{Basic galaxy data}

\begin{tabular}{lccccccrc}
\hline
Galaxy & Morph. & Dist & $b/a$ & $\mu_{0,B}$ & $\log(h_B)$ & $\mu_{0,R}$ & $\log(h_R)$ & $W_{20}$ \\
Name   &type (T)& (Mpc)&       & mag arcsec$^{-2}$ & (kpc) & mag arcsec$^{-2}$ & (kpc) & km s$^{-1}$ \\
 \hline
UGC~00242 & 7   & 56.9 & 0.77  & ---               & ---   & 19.781            & 0.529 & 271 \\ 
UGC~00334 & 9   & 60.2 & 0.77  & 23.163            & 0.780 & 22.200            & 0.774 & 124 \\ 
UGC~00438 & 5   & 58.7 & 0.78  & 20.459            & 0.600 & 19.289            & 0.591 & 372 \\ 
UGC~00490 & 5   & 59.0 & 0.71  & 21.461            & 0.721 & 19.724            & 0.606 & 360 \\ 
UGC~00508 & 2   & 60.7 & 0.89  & 21.422            & 0.865 & 19.686            & 0.797 & 488 \\ 
UGC~00628 & 9   & 70.6 & 0.63  & 22.670            & 0.676 & 21.676            & 0.653 & 228 \\ 
UGC~01305 & 4   & 35.7 & 0.81  & 21.593            & 0.705 & 20.206            & 0.703 & 329 \\ 
UGC~01551 & 8   & 36.0 & 0.89  & 22.147            & 0.653 & 21.008            & 0.624 & 132 \\ 
UGC~01577 & 4   & 69.1 & 0.79  & 21.620            & 0.759 & 20.232            & 0.749 & 287 \\ 
UGC~01719 & 3   &107.6 & 0.74  & 22.163            & 1.029 & 20.688            & 0.982 & 505 \\ 
UGC~02064 & 4   & 56.5 & 0.70  & 21.873            & 0.741 & 20.517            & 0.703 & 263 \\ 
UGC~02124 & 0   & 35.7 & 0.88  & 22.127            & 0.569 & 20.637            & 0.608 & 181 \\ 
UGC~02368 & 3   & 48.0 & 0.92  & 21.142            & 0.568 & 19.822            & 0.512 & 385 \\ 
UGC~02595 & 4   & 78.3 & 0.78  & ---               & ---   & 20.104            & 0.815 & 276 \\ 
UGC~03080 & 5   & 49.5 & 0.91  & 21.825            & 0.661 & 20.397            & 0.570 & 170 \\ 
UGC~03140 & 5   & 63.8 & 0.94  & 20.690            & 0.616 & ---               & ---   & 145 \\ 
UGC~04126 & 3   & 70.2 & 0.85  & 21.436            & 0.820 & 20.024            & 0.764 & 295 \\ 
UGC~04256 & 5   & 75.3 & 0.86  & 21.084            & 0.856 & ---               & ---   & 205 \\ 
UGC~04308 & 5   & 54.1 & 0.77  & 21.191            & 0.701 & 20.041            & 0.665 & 263 \\ 
UGC~04368 & 6   & 58.1 & 0.80  & 21.387            & 0.681 & 20.263            & 0.654 & 248 \\ 
UGC~04375 & 5   & 34.5 & 0.66  & 21.190            & 0.516 & 19.990            & 0.476 & 295 \\ 
UGC~04422 & 4   & 66.1 & 0.82  & 21.978            & 1.004 & 20.657            & 0.968 & 376 \\ 
UGC~04458 & 1   & 69.5 & 0.87  & 22.784            & 0.978 & 21.239            & 0.940 & 285 \\ 
UGC~05103 & 3   & 57.3 & 0.61  & 20.494            & 0.673 & 19.492            & 0.616 & 398 \\ 
UGC~05303 & 5   & 24.8 & 0.59  & 21.207            & 0.592 & 20.069            & 0.547 & 294 \\ 
UGC~05510 & 4   & 24.9 & 0.81  & 20.609            & 0.385 & 19.577            & 0.348 & 187 \\ 
UGC~05554 & 1   & 23.9 & 0.69  & 21.022            & 0.341 & 19.654            & 0.298 & 284 \\ 
UGC~05633 & 8   & 25.9 & 0.63  & 22.967            & 0.523 & 21.978            & 0.545 & 177 \\ 
UGC~05842 & 6   & 22.4 & 0.85  & 21.815            & 0.743 & 20.617            & 0.602 & 166 \\ 
UGC~06077 & 3   & 26.9 & 0.82  & 20.959            & 0.371 & 19.738            & 0.348 & 142 \\ 
UGC~06277 & 5   & 15.7 & 0.88  & 20.890            & 0.333 & 19.730            & 0.284 & 134 \\ 
UGC~06445 & 4   & 24.5 & 0.84  & 20.562            & 0.294 & 19.159            & 0.228 & 184 \\ 
UGC~06453 & 4   & 23.5 & 0.75  & 20.328            & 0.279 & ---               & ---   & 241 \\ 
UGC~06460 & 4   & 23.4 & 0.75  & 20.609            & 0.474 & 19.569            & 0.450 & 202 \\ 
UGC~06693 & 4   & 99.5 & 0.85  & 21.571            & 0.947 & 20.273            & 0.894 & 189 \\ 
UGC~06746 & 0   & 99.2 & 0.76  & 21.406            & 0.921 & 20.355            & 0.959 & 551 \\ 
UGC~06754 & 3   &101.3 & 0.89  & 23.140            & 1.332 & 21.068            & 1.103 & 224 \\ 
UGC~07169 & 5   & 36.3 & 0.81  & 19.827            & 0.343 & 18.613            & 0.263 & 227 \\ 
UGC~07315 & 4   & 09.2 & 0.67  & 20.037            & -0.17 & 18.878           & -0.164 & 250 \\ 
UGC~07450 & 4   & 29.0 & 0.89  & 21.038            & 0.935 & 19.782            & 0.912 & 268 \\ 
UGC~07523 & 3   & 12.3 & 0.81  & 21.143            & 0.279 & 19.869            & 0.241 & 178 \\ 
UGC~07594 & 2   & 33.9 & 0.69  & 21.305            & 0.970 & 20.063            & 0.963 & 306 \\ 
UGC~07901 & 5   & 08.8 & 0.67  & 19.746            & -0.060 & 18.377          & -0.142 & 384 \\ 
UGC~08289 & 4   & 53.2 & 0.89  & 21.711            & 0.940 & 20.244            & 0.781 & 243 \\ 
UGC~08865 & 2   & 38.6 & 0.76  & 21.753            & 0.733 & 20.666            & 0.730 & 302 \\ 
UGC~09061 & 4   & 79.9 & 0.91  & 22.547            & 1.360 & 21.174            & 1.274 & 195 \\ 
UGC~09481 & 4   & 56.4 & 0.74  & 21.270            & 0.688 & 20.122            & 0.650 & 243 \\ 
UGC~09926 & 5   & 30.8 & 0.71  & 19.945            & 0.421 & ---               & ---   & 351 \\ 
UGC~09943 & 5   & 30.7 & 0.64  & 20.207            & 0.451 & 19.284            & 0.460 & 326 \\ 
UGC~10083 & 2   & 28.7 & 0.82  & 20.791            & 0.391 & 19.428            & 0.335 & 178 \\ 
UGC~10445 & 6   & 15.1 & 0.78  & 21.994            & 0.280 & 21.040            & 0.173 & 159 \\ 
UGC~10584 & 5   & 73.0 & 0.88  & 21.837            & 0.915 & 20.539            & 0.852 & 261 \\ 
UGC~11628 & 2   & 55.6 & 0.70  & 22.467            & 1.124 & 20.646            & 0.952 & 503 \\ 
UGC~11708 & 5   & 54.6 & 0.81  & 21.249            & 0.661 & 20.039            & 0.663 & 297 \\ 
UGC~11872 & 3   & 15.7 & 0.74  & 21.015            & 0.222 & 19.032            & 0.129 & 312 \\ 
UGC~12151 &10   & 23.0 & 0.87  & 22.822            & 0.414 & 21.981            & 0.361 & 180 \\ 
UGC~12343 & 5   & 31.0 & 0.77  & 21.023            & 0.767 & 19.800            & 0.766 & 376 \\ 
UGC~12379 & 4   & 80.6 & 0.91  & 21.792            & 0.886 & 20.326            & 0.893 & 308 \\ 
UGC~12391 & 5   & 62.9 & 0.89  & 21.312            & 0.685 & 20.071            & 0.665 & 216 \\ 
UGC~12511 & 6   & 45.9 & 0.82  & 22.282            & 0.712 & 20.685            & 0.567 & 278 \\ 
UGC~12614 & 5   & 45.0 & 0.74  & 20.816            & 0.657 & 19.609            & 0.636 & 315 \\ 
UGC~12638 & 5   & 72.7 & 0.81  & 21.913            & 0.861 & 20.615            & 0.830 & 303 \\ 
UGC~12654 & 4   & 52.1 & 0.82  & 21.519            & 0.684 & 20.340            & 0.655 & 237 \\ 
\end{tabular}										    
\end{minipage}										    
\end{table*}										    
											    
\clearpage										    
\clearpage										    
											    
\setcounter{table}{0}									    
\begin{table*}										    
\centering
\begin{minipage}{140mm}
\caption{-- {\it continued}}
\medskip
\begin{tabular}{lccccccrc}
\hline
Galaxy & Morph. & Dist & $b/a$ & $\mu_{0,B}$ & $\log(h_B)$ & $\mu_{0,R}$ & $\log(h_R)$ & $W_{20}$ \\
Name   &type (T)& (Mpc)&       & mag arcsec$^{-2}$ & (kpc) & mag arcsec$^{-2}$ & (kpc) & km s$^{-1}$ \\
 \hline
UGC~12732 & 9   & 11.4 & 0.89  & ---               & ---   & 21.515            & 0.087 & 127 \\ 
UGC~12754 & 6   & 11.5 & 0.70  & 21.231            & 0.370 & 20.207            & 0.321 & 206 \\ 
UGC~12776 & 3   & 64.1 & 0.88  & 23.305            & 1.355 & 21.834            & 1.149 & 256 \\ 
UGC~12808 & 3   & 54.4 & 0.93  & 19.839            & 0.503 & 18.659            & 0.504 & 362 \\ 
          &     &      &       &                   &       &                   &       &     \\ 
F561-1    & 9   & 70.1 & 0.91  & 22.738            & 0.478 & 21.843            & 0.437 & 069 \\ 
F563-1    & 9   & 53.9 & 0.91  & 23.716            & 0.730 & 22.602            & 0.628 & 115 \\ 
F563-V1   &10   & 59.5 & 0.500 & 23.669            & 0.450 & 22.358            & 0.432 & 097 \\ 
F563-V2   &10   & 64.5 & 0.88  & 22.061            & 0.368 & ---               & ---   & 132 \\ 
F564-v3   &10   & 05.9 & 0.87  & 23.824            & -0.519 & 22.890          & -0.588 & 065 \\ 
F565-V2   &10   & 56.7 & 0.50  & 24.065            & 0.463 & 23.394            & 0.438 & 126 \\ 
F567-2    & 9   & 83.5 & 0.94  & 23.995            & 0.692 & 22.936            & 0.577 & 074 \\ 
F568-1    & 5   & 94.1 & 0.90  & 23.617            & 0.775 & 22.619            & 0.710 & 150 \\ 
F568-3    & 7   & 86.1 & 0.77  & 22.571            & 0.575 & 21.611            & 0.528 & 180 \\ 
F568-V1   & 7   & 84.5 & 0.77  & 22.741            & 0.484 & 21.942            & 0.466 & 176 \\ 
F571-5    & 9   & 65.1 & 0.77  & 23.362            & 0.496 & 22.647            & 0.448 & 179 \\ 
F571-V1   & 7   & 84.2 & 0.82  & 23.590            & 0.511 & 22.637            & 0.472 & 084 \\ 
F5741-    & 7   & 99.2 & 0.42  & ---               & ---   & 22.386            & 0.749 & 223 \\ 
F574-2    & 9   & 91.9 & 0.88  & 24.102            & 0.754 & 23.027            & 0.635 & 068 \\ 
F577-V1   & 7   &103.7 & 0.82  & 23.787            & 0.799 & 23.046            & 0.718 & 094 \\ 
F583-1    & 9   & 35.1 & 0.45  & 23.012            & 0.318 & 22.234            & 0.275 & 183 \\ 
UGC~1230  & 9   & 50.4 & 0.94  & 23.208            & 0.677 & 22.333            & 0.609 & 115 \\ 
UGC~128   & 8   & 58.9 & 0.55  & 23.405            & 0.864 & 22.488            & 0.764 & 214 \\ 
UGC~6614  & 1   & 92.1 & 0.59  & ---               & ---   & 22.567            & 1.114 & 262 \\ 
1150280   & 5   & 88.2 & 0.62  & 20.772            & 0.657 & 19.639            & 0.640 & 190 \\
1530170   & 4   & 88.3 & 0.71  & ---               & ---   & 21.369            & 0.941 & 198 \\
2520100   & 3   &130.9 & 0.76  & 23.894            & 1.406 & 21.106            & 0.929 & 434 \\    
0350110   & 1   & 86.1 & 0.77  & 20.829            & 0.690 & 19.529            & 0.648 & 285 \\  
3740090   & 1   & 41.8 & 0.81  & 22.704            & 0.428 & 21.145            & 0.374 & 076 \\
4220090   & 7   & 65.9 & 0.91  & 22.901            & 0.717 & 21.796            & 0.589 & 078 \\
4250180   & 7   & 95.1 & 0.84  & 23.378            & 0.945 & 21.932            & 0.821 & 101 \\
4990110   & 7   & 41.0 & 0.83  & 22.738            & 0.423 & 21.995            & 0.498 & 090 \\
0050050   & 8   & 77.2 & 0.92  & 22.453            & 0.705 & 21.454            & 0.621 & 132 \\
0540240   & 2   &198.7 & 0.84  & 23.555            & 1.111 & 22.289            & 1.012 & 399 \\
5520190   & 2   &159.5 & 0.62  & 22.488            & 1.115 & 20.810            & 0.942 & 459 \\
0590090   & 8   & 18.4 & 0.88  & 21.847            & 0.283 & 20.812            & 0.262 & 053 \\
1220040   & 1   &127.0 & 0.63  & 23.272            & 1.227 & 21.213            & 0.882 & 101 \\
\hline 
\end{tabular}
\medskip
Column 1 gives the galaxies identification.  The HSB sample from de Jong 
\& van der Kruit (1994) are designated according to their UGC number; the 
LSB galaxies from de Blok et al.\ (1995) which are labelled F (except for
3 galaxies labelled with a UGC number) are from 
the lists of Schombert et al.\ (1992); the LSB galaxies from 
Beijersbergen et al.\ (1999) are denoted simply with a numeric code 
which refers to the ESO-LV number from the catalog of Lauberts \& 
Valentijn (1989).  The galaxy type, or numerical stage index T 
(de Vaucouleurs et al.\ 1991), is given column 2. 
Column 3 gives the galaxy distance in Mpc (assuming 
$H_0=75$ km s$^{-1}$ Mpc$^{-1}$) after correcting the 
observed redshift (courtesy of NED) for Galactic rotation, 
Virgo-infall and infall towards the `Great Attractor' using the 
model and code from Burstein et al.\ (1989).  Column 4 gives the 
($R$-band) axis-ratio of each disk.  The central $B$-band disk 
surface brightness and disk scale-length are given in Columns 5 and 6.
The surface brightness term has been corrected for the various effects
described in Section 2 of the text.  Column 7 and 8 are the $R$-band  
equivalent of Column 5 and 6.  Column 9 gives the observed linewidth 
at 20 per cent of the peak HI intensity.
\end{minipage}
\end{table*}

Redshifts for ESO-LV 0350110 and ESO-LV 0050050, from the 
bulge-dominated sample of LSB galaxies, 
were obtained from the Parkes telescope HIPASS database
(Barnes et al.\ 2001).  Redshifts for the other galaxies 
have been taken from the NASA/IPAC Extragalactic Database (NED).  
The heliocentric redshifts were uniformly 
corrected for Galactic rotation, Virgo-infall, infall toward the 
`Great Attractor', and the `local anomaly' using the model from 
Burstein et al.\ (1989).  The disk scale-lengths were then converted 
from arcseconds to kpc using a Hubble constant 
$H_0=75$ km s$^{-1}$ Mpc$^{-1}$.  

In passing, 
it is noted that the sample of galaxies has been taken from the field,
rather than from a single cluster, and so one can expect to have a greater 
degree of scatter in the relationship between $\log v$, $\mu_0$, and 
$\log h$ than might otherwise be attained.  This is because peculiar 
velocity flows leading to distance errors will introduce errors into 
the estimates of the scale-lengths.  Additionally, 
like most studies to date, $W_{20}$ has been used to derive $v_{\rm rot}$ 
rather than using $v_{max}$ or $v_{\rm flat}$ which are not available. 
It is acknowledged that the available kinematic data may therefore not be 
the optimal estimate to the rotational velocity, although McGaugh \& 
de Blok (1998) state that $W_{20}$ gives a good approximation to 
$v_{\rm flat}$.  


\section{The relation between $\log \lowercase{v}$, $\mu_0$ and 
$\log \lowercase{h}$, and implications for the $M/L$ ratio}

From the balance of forces between centripetal acceleration and 
gravity, one has the dynamical mass $M(r)=v^2r/G$, where $G$ is the 
gravitational constant.  Combining the previous expressions for 
$M$ and $L$ (section 1) gives 
\begin{equation} 
\frac{M}{L} \propto \frac{v^2r}{I_0h^2}, 
\end{equation}
where $M$ and $v$ are measured at the same radius $r$.  
A common practice is to take $r$ as some multiple of $h$ (e.g.\ 4$h$, 
McGaugh \& de Blok 1998) and assume the estimate of $v$ derived from 
$W_{20}$ is a good estimate of not only $v_{\rm flat}$ 
(McGaugh et al.\ 2000) but also of the rotational velocity at 4$h$. 
Doing this, one arrives at the virial theorem 
\begin{equation}
v^2 \propto (M/L)I_0h. 
\label{eqx}
\end{equation}

Next, from the measured data of many galaxies, one can construct -- given 
it exists -- the `plane'
\begin{equation}
v\propto I_0^x h^y,
\label{eqy}
\end{equation}
where $x$ and $y$ are the optimal exponents dictating the relationship
between the disk's structural and dynamical terms. 
Combining equations~\ref{eqx} and ~\ref{eqy} to eliminate 
the velocity term, one quickly arrives at the relation
\begin{equation}
-2.5\log \frac{M}{L} \propto (2x-1)\mu_0 - 2.5(2y-1)\log h.
\label{eqMoL}
\end{equation}
Subsequently, lines of constant $M/L$ will have a slope of
$2.5(2y$$-$$1)/(2x$$-$$1)$ in the $\mu_0$--$\log h$ diagram 
(e.g.\ Graham \& de Blok 2001). 
One can similarly derive that lines of constant velocity have a 
slope of $2.5y/x$.

If the value of $v_{\rm rot}$ derived from $W_{20}$, or $v_{\rm flat}$, 
is an appropriate velocity to use at some constant multiple of $h$
(e.g.\ 4$h$), and galaxy disks are virialised, then the $M/L$ ratio 
derived in equation~\ref{eqMoL} should be proportional to the $M/L$ 
ratio derived from 
\begin{equation}
M/L\mid _{4h} \propto v^24h/(I_0h^2). 
\label{MLat4}
\end{equation} 
I shall refer to this as the `kinematical' or 'dynamical' 
mass-to-light ratio, and that in equation 4 as the 
`photometric' mass-to-light ratio. 

What $M/L$ ratio does one really want to measure? 
Lets start by looking at the different ways which can be used
to measure $L$. 
The flux within some limiting isophotal radius is clearly 
inappropriate.  This is strikingly evident in the case of LSB galaxies 
whose central surface brightness values may be fainter than the 
limiting isophotes used to measure the HSB galaxies.  In such 
an instance, the flux measurement of the LSB galaxy would be zero. 
The use of the outer 
most observed radius is also not ideal because it depends on 
observational details such as exposure times, telescope aperture, and 
what signal-to-noise one is willing to accept; that is, the outer 
most observed radius is not an intrinsic quantity of a galaxy.  
Another way to estimate the luminosity of galaxies is to measure 
their flux within some fixed aperture size, say 20 kpc. 
Although this is indeed a standard and consistent approach, it is 
again not a favourable one.  
The reason is because, depending on $\mu_0$ and $h$, {\it this metric 
will measure a different fraction of the total galaxy luminosity.} 
Given that exponential disks are structurally homologous, scaling with 
the central intensity and scale-length, one can, and indeed should, 
use these quantities to measure the same fraction of disk light 
for all galaxies (for example, 91 per cent of the light from an exponential 
profile which extends to infinity is contained within the inner 
four scale-lengths; 99.1 per cent within the inner 4 effective radii).  
Therefore, some constant multiple of $I_0h^2$ will represent the 
same fraction of the total extrapolated disk light for all disks. 
Disk truncation in real galaxies at 4 or more scale-lengths will 
introduce less than a 10 per cent error to this approach. 

What about the estimate for the mass $M$?  One could also measure the 
galaxy mass within 20 kpc, or within a fixed number of disk 
scale-lengths.  But is this really what we want?  If the total (stars
plus gas plus dark matter) mass profile does not scale with the 
luminous disk scale-length, then 
estimates for the galaxy mass within a fixed number of disk scale-lengths 
will not correspond to equal fractions of the mass distribution. 
Therefore, a more fundamental quantity might be the mass within 
some constant multiple of the core-radii of the {\it total} mass profile, 
defined to be the radius where the density has dropped by a factor of 
two from its central value. 
If the mass profiles are homologous, then the mass would simply scale
with some measure of velocity and some measure of radius.  
The presence of the 
flat rotation curves suggests a velocity scale we can use, namely 
$v_{\rm flat}$.  This then leaves the radial scale-size 
to be determined, and one typically 
assumes that the scale-size `$a$' of the {\it total} mass profile, scales 
with the disk scale-length (McGaugh \& de Blok 1998)\footnote{Alternative 
scales could be the velocity $v_0$ at the radius $r_0$ from the density 
profiles used in Kravtsov et al.\ (1998).}.  If they do not, 
then one should add the term $-2.5\log(a/h)$ to the right hand side of 
equation~\ref{eqMoL}.  

Whether or not galaxy rotation curves are homologous appears to be 
something of an unresolved question (van den Bosch et al.\ 2000). 
Swaters, Madore \& Trewhella (2000; their figure 4) 
%
%
present H$_{\alpha}$
rotation curves for a small sample of LSB and HSB galaxies which 
appear to overlay each other exactly (once normalised with $v_{\rm flat}$
and $h$).  This contradicts the work of de Blok \& McGaugh (1997) and 
suggests that $a/h$ may be constant.  
Most recently, fitting pseudo-isothermal mass models (in preference
to Navarro, Frenk \& White (1996) CDM halos), de Blok, McGaugh \& Rubin 
(2001; their figure 10) studied a sample of well-resolved LSB 
galaxy rotation curves and found no appreciable evidence for any 
correlation between $a/h$ and the maximum rotational velocity, or 
equivalently luminosity.  

In what follows, the additional term $a/h$ will be ignored and the
kinematical $M/L$ ratio shall be that within 4 scale-lengths, as 
defined in equation~\ref{MLat4}.  

If the TF relation is indeed the optimal representation of
disk galaxies, then one must find from equation~\ref{eqy} 
that $y$$=$$2x$ and $v \propto L^x$.  
The exact slope of the TF relation is important because it tells 
us the $M/L$ dependency on the other galaxy parameters.   
For example, if $L \propto v^2$ ($x$$=$0.5, $y$$=$1.0) then 
$(M/L) \propto h$ and is independent of surface brightness.\footnote{This 
result actually holds for any value of $y$ as long as $x$$=$0.5.}  
However, if $L \propto v^4$ ($x$$=$0.25, $y$$=$0.5) then 
$(M/L) \propto 1/\sqrt{I_0}$, independent of galaxy size. 

In the original paper of Tully \& Fisher (1977), they found 
a slope of 2.5$\pm$0.3 using photographic magnitudes and the 
velocity line width at 20 per cent of the maximum intensity.
More recently, 
Zwaan et al.\ (1995) obtained a $B$-band slope of 2.6, implying that 
$\log(M/L_B)\propto (0.09\mu_0+0.54\log h)$\footnote{n.b.\ Along lines of 
constant luminosity in the $\mu_0$--$\log h$ diagram, 
if $\log(M/L)\propto 0.1\mu_0 + 0.5\log h$ then it implies that
both $\mu_0$ and $\log h$ contribute equally to changes in $M/L$.}
This result is not at odds with the conclusion of Zwaan et al.\ (1995)
that $M/L \propto I_0^{-1/2}$, because their statement only holds 
valid, as they state, for galaxies of a fixed/constant luminosity. 
This point has, however, not always been 
appreciated in the literature. 

Let us consider the case where $L\propto v^2$ and 
$\log(M/L)\propto \log h$, independent of $\mu_0$.  
If one was to sample different galaxies having the same $h$ but 
different $\mu_0$, $\log(M/L)$ would show no dependency 
at all on $\mu_0$.  If one was to then sample a different set of 
galaxies having equal 
luminosity, but with different $\mu_0$ and $h$, a plot of $\log(M/L)$ 
versus $\mu_0$ would of course show a strong correlation 
This is because of the relation between $\mu_0$ and h at constant 
$L$.  One must therefore be careful of misleading results of this 
kind.  Completely independently of any observational results, 
one can see this degeneracy at constant $L$ by 
dividing the relation $M\propto v^2h$ by $L$ to give 
$(M/L)\propto v^2h/L$. 
If the TF relation holds true with any slope at all, 
then one has that $(M/L)\propto h$ for any fixed $L$.  
{\it Of course, for different values of $L$ the constant of 
proportionality between $M/L$ and $h$ is different (unless 
$L\propto v^2$) and 
so one cannot use this expression to compare the $M/L$ ratios 
between galaxies of different luminosity, only between galaxies of  
the same luminosity.}  Alternatively, Zwaan et al.\ (1995) showed that 
$I_0(M/L)^2$ is constant at any fixed linewidth (or, from the TF
relation, any fixed luminosity); therefore, $(M/L)\propto 1/\sqrt{I_0}$.
However, if $I_0h^2$ is constant, one again has that
$(M/L)\propto h$.  Identically, {\it one cannot use the relation 
$(M/L)\propto 1/\sqrt{I_0}$ to compare the $M/L$ ratio of 
galaxies at different luminosities (unless $L\propto v^4$).} 
Only for galaxies having the same fixed luminosity can these 
expressions be used to indicate differences in the $M/L$ ratio. 
How the $M/L$ ratio of galaxies {\it with different luminosities} 
varies (simultaneously) with scale-length and the central disk 
intensity is what this study shall investigate.

\section{Data analysis and results}

A Principle Component Analysis was performed on the data, revealing 
that 95 per cent ($B$-band) and 94 per cent ($R$-band) of the variance 
within the 3-dimensional space of variables 
($\mu_0$, $\log h$, $\log v$) resides within a two-dimensional 
plane\footnote{The principle eigenvector accounted for 49 per cent
($B$-band) and 55 per cent ($R$-band) of the variance.}.
From this it is concluded that it is indeed appropriate to represent the 
data with a two-dimensional plane.   In so doing, five galaxies 
(3 LSB and 2 HSB) from 
the final sample of 94 have been identified as potential outliers -- 
one in particular (ESO-LV 1220040).  The following analysis has therefore 
been performed excluding ESO-LV 1220040, and then performed again excluding 
all five galaxies, which, in a number of tests, consistently lied 
away from the main data.  These points are shown in 
Figure~\ref{fig2} -- identified as lying more than a factor of 2 in 
$v_{\rm rot}$ from the best-fitting line to the data sample.  
Two\footnote{An additional 10 galaxies have inclinations $<$25$^\circ$, 
but all closely follow the relation defined by the other galaxies.} 
of these galaxies have an inclination of $\sim$25$^\circ$, and another 
is the closest galaxy from the HSB sample -- having a redshift of only 
800 km s$^{-1}$.  The above percentages increased by 1 per cent, to 
96 per cent ($B$-band) 
and 95 per cent ($R$-band), when all five outlying data points were excluded.  

\begin{figure}
\centerline{\psfig{figure=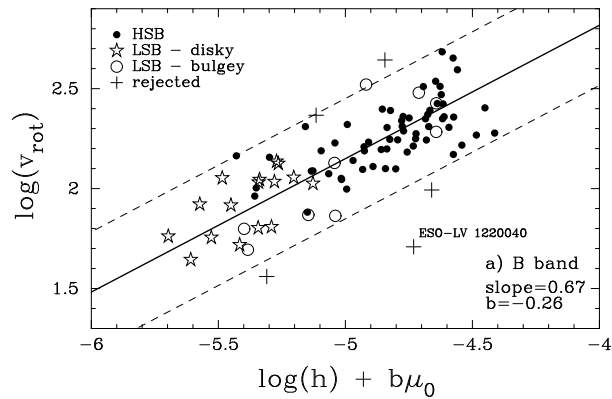,width=8.0cm,angle=-90}}
\caption{
Correlation between the dynamical and structural properties of spiral
galaxy disks. 
The `mixing' value $b$ (see, e.g., Djorgovski \& Davis 1987) is 
chosen such that it yields the highest linear correlation 
coefficient between $\log v_{\rm rot}$ and $(\log h + b \mu_0)$. 
The solid line represents the best fit from an orthogonal regression
analysis, and the dashed lines mark the boundary which is a factor of 2
away in $v_{\rm rot}$ from the line of regression.  The different symbols 
denote from which galaxy sample the data are from: filled circles 
represent galaxies 
from the sample of de Jong \& van der Kruit (1994), stars are for 
galaxies from de Blok et al.\ (1995), open circles are for galaxies from 
Beijersbergen et al.\ (1999), and the plus signs denote the five outlying 
points which have been rejected from the analysis. 
The perpendicular scatter in both panels a) $B$-band and b) $R$-band
is 0.11 dex.  
}
\label{fig2}
\end{figure}

\begin{figure}
\centerline{\psfig{figure=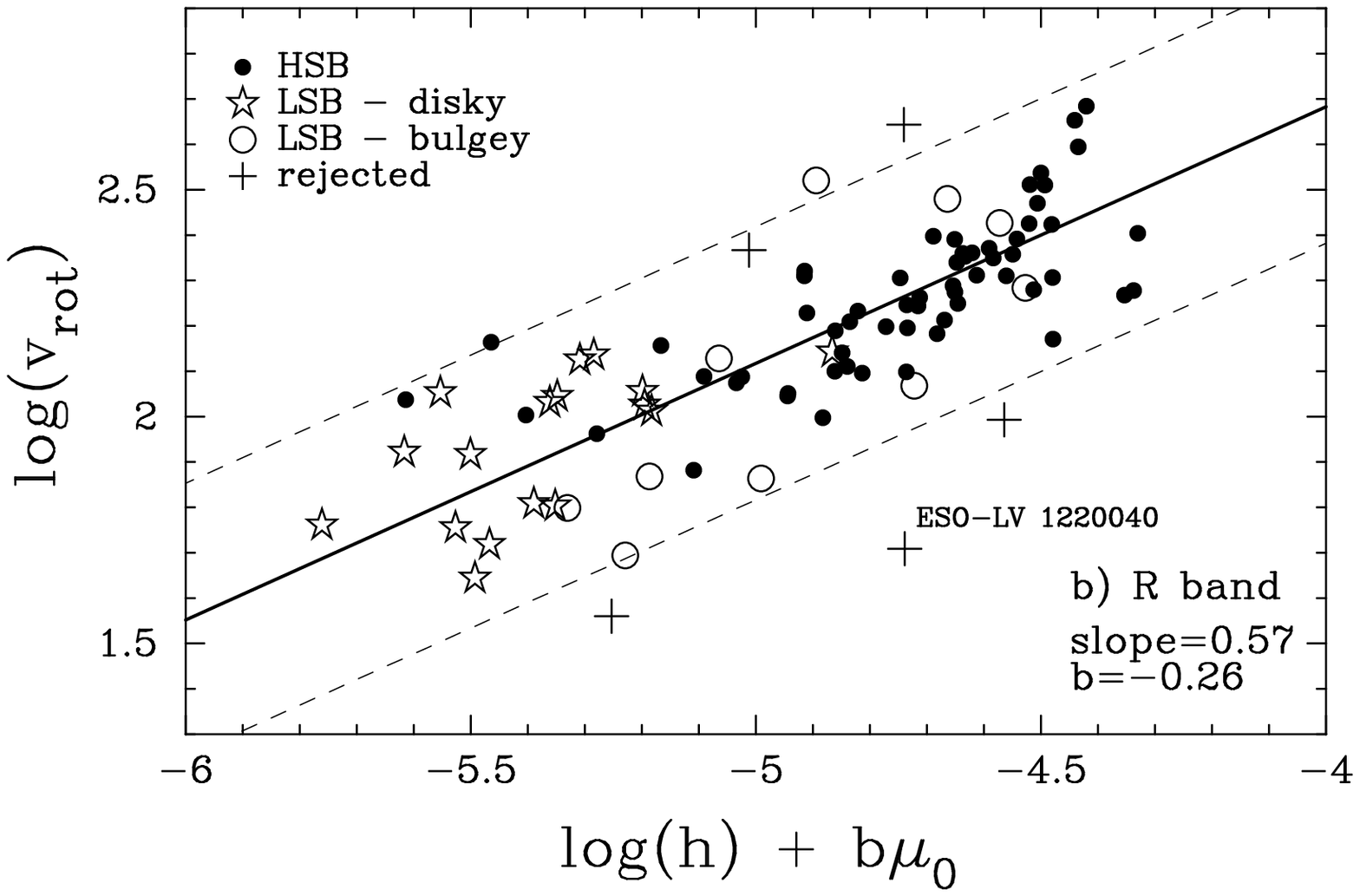,width=8.0cm,angle=-90}}
\contcaption{ }
\end{figure}


To construct the best-fitting plane, a three-dimensional
bisector method of linear-regression was used (Graham \& 
Colless 1997).  The advantages and applicability in treating all the 
variables equally in this manner, as opposed to a simple linear 
regression of one variable on the others, is elucidated in Isobe 
et al.\ (1990) and Feigelson \& Babu (1992). 
Basically, one wants to treat all the data/variables in an equivalent 
manner when performing a comparison with theory.  A simple ordinary
linear regression (OLS) of one variable on the other two variables would 
not achieve this goal, and would produce different results depending 
on which 
variable one regressed upon the others.  Of those methods which do 
provide a symmetrical treatment of the data, the bisector method 
provides the smallest variance, or uncertainty, to the 
estimated values of the fitted plane. 

Performing this regression, the exponents $x$ and $y$ from 
equation~\ref{eqy} were computed to be 0.53$\pm$0.05 and 
0.79$\pm$0.09 ($B$-band) and 0.45$\pm$0.04 and 0.71$\pm$0.09 ($R$-band). 
Therefore, $(M/L_B)\propto I_0^{0.06\pm0.10}h^{0.58\pm0.18}$ ($B$-band) 
and $(M/L_R)\propto I_0^{-0.10\pm0.08}h^{0.42\pm0.18}$ ($R$-band).  
Lines of constant velocity subsequently have a slope of 
$2.5y/x=3.7$$\pm$0.5 ($B$-band) and 3.9$\pm$0.6 ($R$-band) 
in the $\mu_0$--$\log h$ diagram.  
A plot of $\log v$ against $\log(I_0^xh^y)$ does, as one would 
hope, have a slope of 1.  This figure is not presented here 
because it is 
attained by simply dividing the $x$-axis in Figure~\ref{fig2} by the 
slope of that data.  Therefore, a scaled version of such a plot is 
already displayed.  Both of the TF relations shown in Figure~\ref{fig3} 
(obtained by setting the `mixing' value $b\equiv -x/(2.5y)=-0.2$) have 
a vertical scatter of 0.14 dex, slightly greater than the value 
0.13 dex (c.f.\ 0.65 mag, Kannappan, Fabricant \& Franx 2002) 
obtained in Figure~\ref{fig2}.

\begin{figure}
\centerline{\psfig{figure=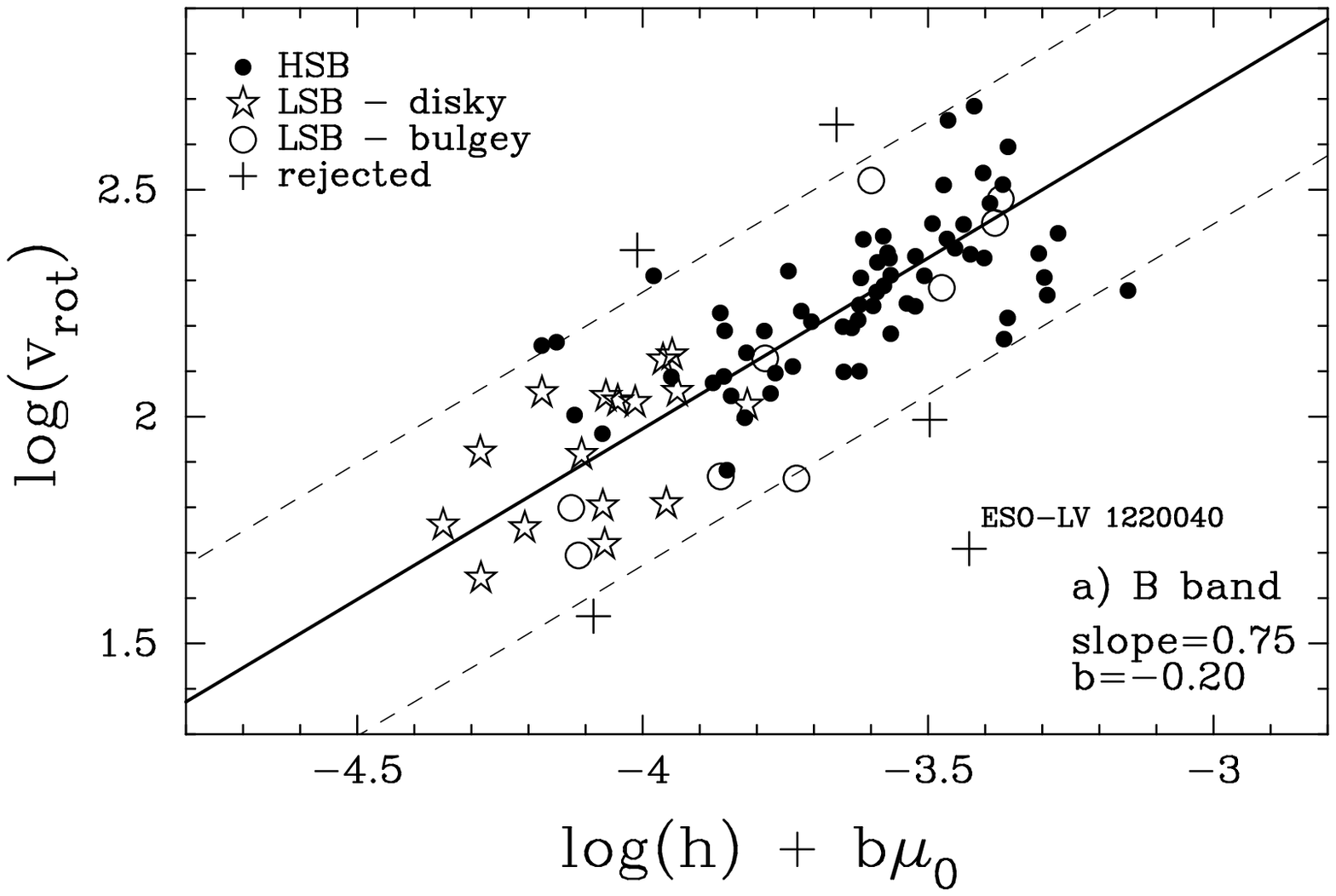,width=8.0cm,angle=-90}}
\caption{Same as Figure~\ref{fig2}a and \ref{fig2}b, 
except that the value of $b$ has been fixed equal to -0.2 
to mimic the TF relation, where $L\propto v^{\alpha}$ and $\alpha$ is
obtained here by dividing 2.0 by the slope of the fit.  
The perpendicular scatter in both panels a) $B$-band and b) $R$-band
is 0.11 dex. 
}
\label{fig3}
\end{figure}

\begin{figure}
\centerline{\psfig{figure=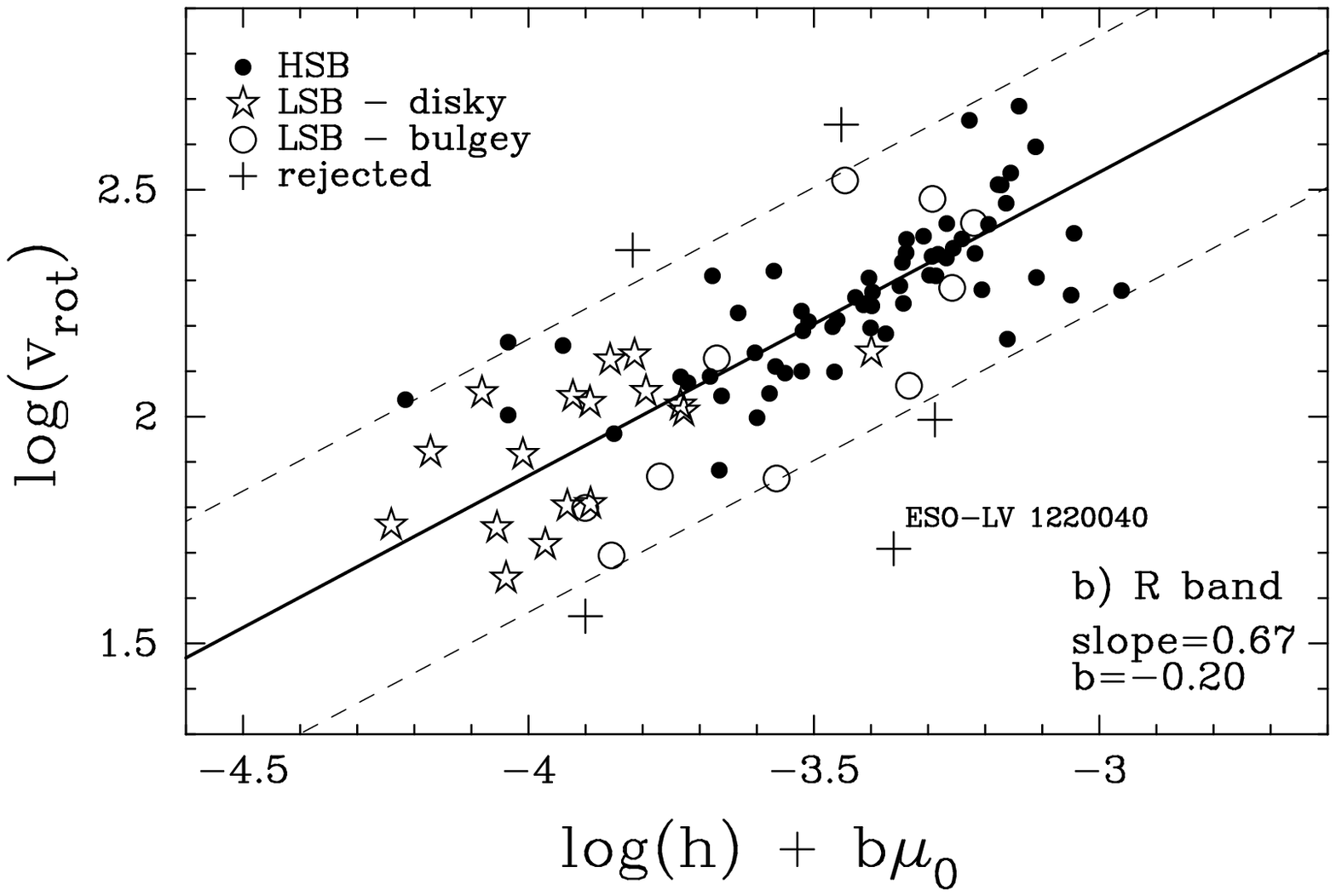,width=8.0cm,angle=-90}}
\contcaption{ }
\end{figure}

The above analysis was repeated, removing those {\it additional} 
four data points marked as outliers in Figure~\ref{fig2}. 
This time $x$$=$0.50$\pm$0.05 and $y$$=$0.77$\pm$0.07
($B$-band), and $x$$=$0.43$\pm$0.04 and $y$$=$0.69$\pm$0.07 ($R$-band). 
Here, $(M/L_B)\propto I_0^{0.00\pm0.10}h^{0.54\pm0.14}$ ($B$-band) and 
$(M/L_R)\propto I_0^{-0.14\pm0.08}h^{0.38\pm0.14}$ ($R$-band), 
and lines of constant velocity have a slope of 3.9$\pm$0.5 ($B$-band) 
and 4.0$\pm$0.6 ($R$-band) in the $\mu_0$--$\log h$ diagram.  
At least in the $B$-band, the data reveal that the $M/L$ ratio 
is in fact largely independent of surface brightness, and roughly 
varies as $\sqrt h$.  {\it Lower surface brightness does not necessarily 
imply a higher $M/L$ ratio.}  The residuals are shown not to correlate 
with either ($B-R$) colour (Figure~\ref{ellip}), or galaxy ellipticity,  
and hence no obvious additional dependency on inclination is expected. 
Studies which explore the density profiles of dark matter halos 
associated with LSB galaxies should perhaps reconsider the assumption 
that all of these systems are dark matter dominated. 

\begin{figure}
\centerline{\psfig{figure=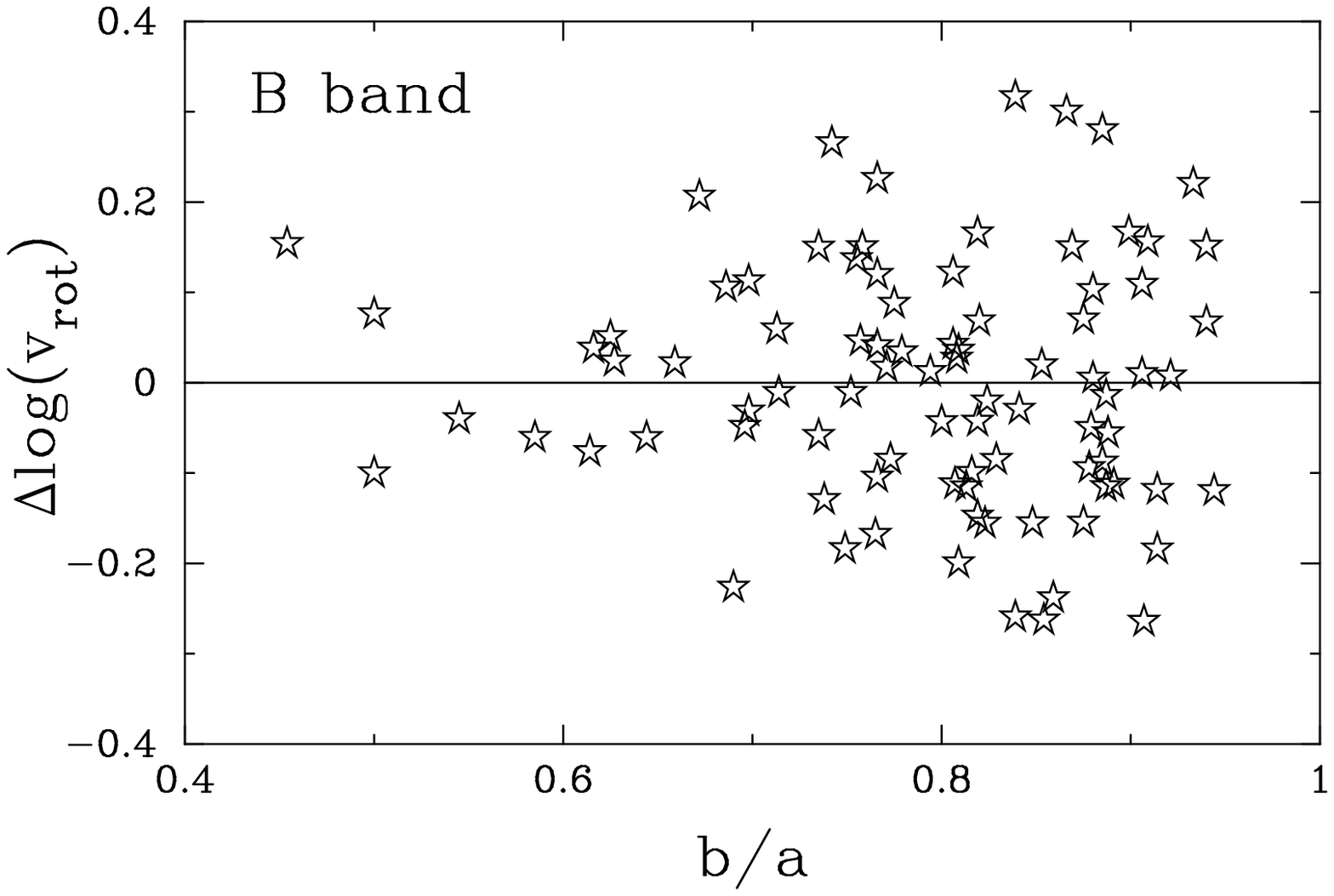,width=8.0cm,angle=-90}}
\caption{The vertical residuals about the best-fitting line in 
Figure~\ref{fig2}a are plotted against a) each galaxies observed 
minor-to-major axis ratio ($b/a$), and b) the $B-R$ colour of each disk.
}
\label{ellip}
\end{figure}

\begin{figure}
\centerline{\psfig{figure=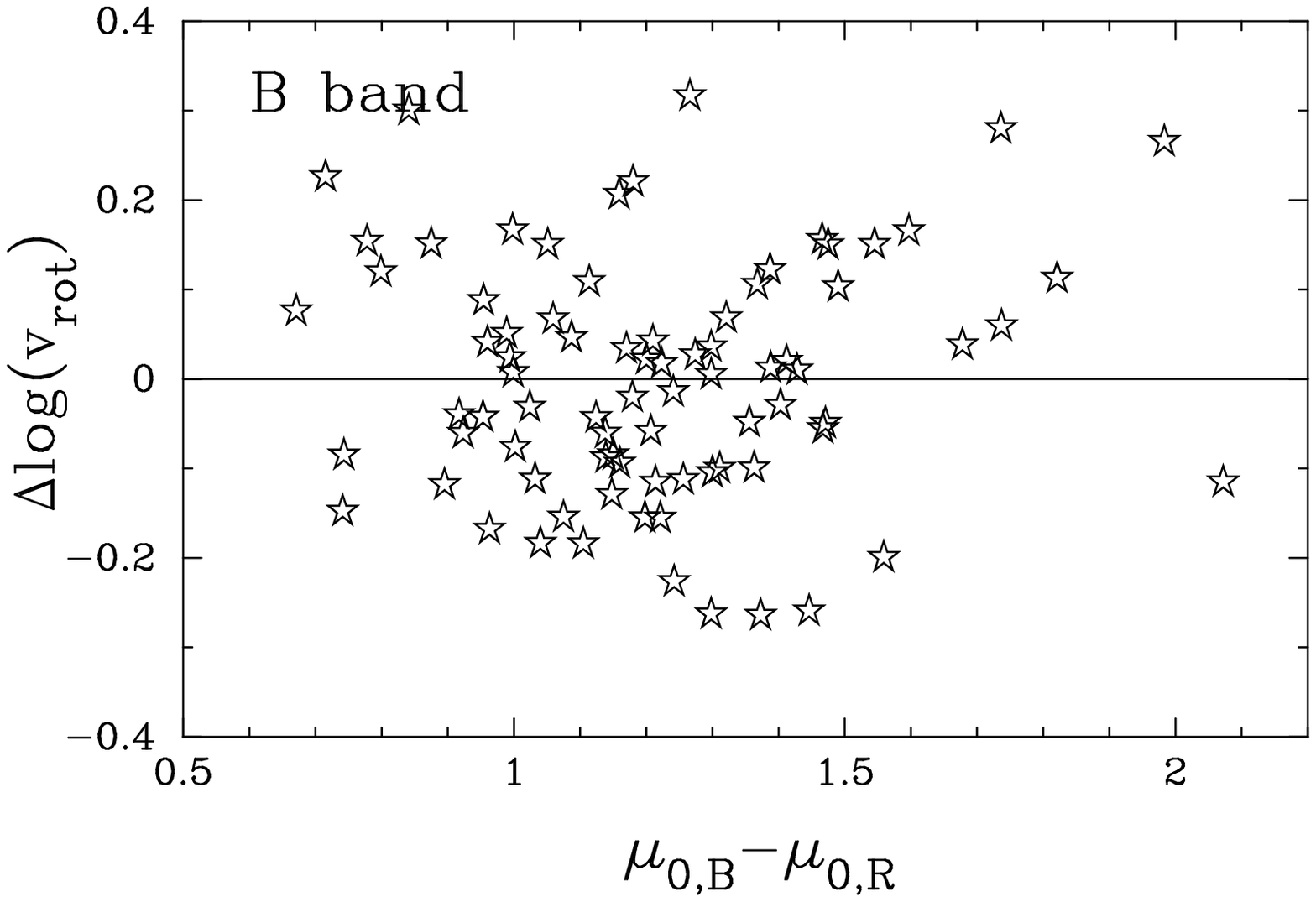,width=8.0cm,angle=-90}}
\contcaption{ }
\end{figure}

\begin{figure}
\centerline{\psfig{figure=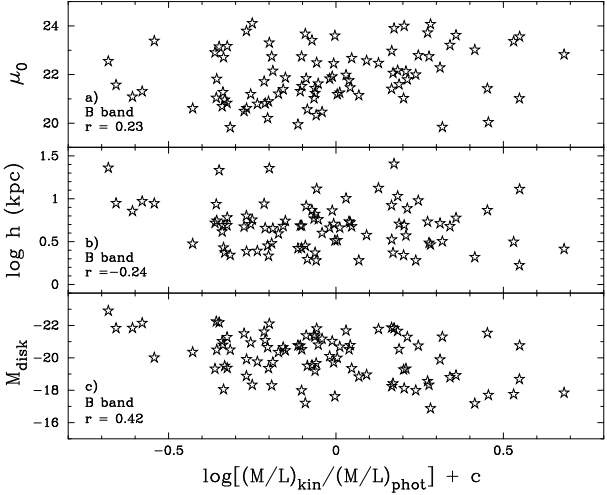,width=8.0cm,angle=-90}}
\caption{
The logarithm of the difference between the mass-to-light ratio 
estimates obtained using equation~\ref{eqMoL} and equation~\ref{MLat4}
have been plotted against a) the central disk surface brightness, 
b) the logarithm of the disk scale-length, and c) the absolute 
magnitude of the disk.  No strong trends exists. 
These plots show the $B$-band data, excluding the 5 outlying points
identified in Figure~\ref{fig2}.  The values for $x$ and $y$ used
in equation~\ref{eqMoL} have been taken to be 0.50 and 0.77. 
Constants of proportionality have been ignored; 
the constant $c$ in the $x$-axis is the same for all galaxies. 
}
\label{figML2} 
\end{figure}

Differences in the mass-to-light ratios estimated from 
equation~\ref{eqMoL} and equation~\ref{MLat4} may signify that 
galaxies are either not virialised or that the mass profiles do 
not scale simply with the luminous exponential disk scale-length 
(section 3).  The logarithm of the ratio of the `kinematic' and 
`photometric' mass-to-light ratios has been 
plotted against the central disk surface brightness, the logarithm of 
the luminous disk scale-length, and the logarithm of the disk luminosity 
in Figure~\ref{figML2}.  No strong correlations exist, although a weak 
correlation (Pearson's $r=0.4$) with absolute magnitude does exist at 
the 99.97 per cent confidence level.  This may be indicating that 
the ratio of the mass profile 
scale-size ($a$) to the ($B$-band) luminous disk scale-length ($h$) 
may correlate weakly with luminosity, in the sense that 
$a/h$ decreases with increasing disk luminosity.

One way to test if the above trend is true would be from an 
analysis of actual rotation curves.  If rotation curves are 
homologous, scaling in amplitude with $v_{\rm flat}$ and radially 
with disk scale-length $h$, then they should all overlap each other
once normalised in this way.  However, because rotation 
curves can rise slowly as they approach $v_{\rm flat}$ (see, for e.g.,
Rubin et al.\ 1985), it is 
difficult to quantify at exactly which radius $v_{\rm flat}$ is reached.
It can therefore be difficult to gauge if rotation curves are homologous. 
It is proposed here that one might be able to use the radius where 
the velocity equals one-half of $v_{\rm flat}$
as a measure of the radial scale for the velocity profile.  Due to the
relative steepness of rotation curves around this value (after they
have been scaled with the disk scale-length), errors in $v_{\rm flat}$
should have a reduced effect on estimates of this radius. 
A plot of this radius, in units of disk scale-length, 
against disk luminosity may reveal if the assumption that the mass 
profiles scale with the luminous disk profiles is valid, or if the slight 
correlation found above is real.

\subsection{Further analysis}

A range of methods have been used in the literature for the task of
constructing a two-dimensional plane to three-dimensional data. 
Given the lack of universal agreement as to which is the best 
approach to take, a second (different) approach is employed here. 
Following Djorgovski \& Davis (1987), the `mixing' value 
$b$ that gave the highest linear correlation coefficient between 
$\log v$ and $(\log h + b \mu_0)$ was computed.  
A line of regression to this two-parameter data set (once the value
of $b$ had been determined) was then fitted; 
in fact, six different lines of regression were fitted using the 
program SLOPES from Feigelson \& Babu (1992).  The results are given 
in Table~\ref{tab1}, and were obtained excluding the 1 outlying 
data point (ESO-LV 1220040).   Through a bootstrap resampling of 
the data an estimate for the error on $b$ was obtained and is also 
given in Table~\ref{tab1}. 
When all five outlying data points were excluded, 
$b$$=$$-0.256\pm0.030$ ($B$-band) and $b$$=$$-0.265\pm0.036$ 
($R$-band), in agreement with the results obtained in Table~\ref{tab1}. 
Again, the analysis and data suggests that in the $B$-band $x\sim 0.5$ 
and the $M/L_B$ ratio is largely independent of surface brightness.  
Once more, only mild inconsistencies with the TF expectation 
value of $b=-0.2$ were found using this alternative technique to 
compute the best-fitting plane.  Within the 2$\sigma$ confidence 
interval, the optimal values of $b$ are consistent with the value -0.20.  
Forcing the value of $b$ to equal -0.20 introduced a 
slight dependency of $M/L$ on the surface brightness term -- as 
evidenced by the departure of $x$ from the value 0.5 in Table~\ref{tab1}. 
The extent of this effect was greater in the redder passband.

\begin{table*}
\begin{minipage}{110mm}
\caption{Regression analysis: Field Galaxies}
\label{tab1}
\begin{tabular}{@{}lccc}
 method  &  intercept  &  slope ($=y$)  &  $x=-2.5\times b\times y$\\[10pt]
$B$-band: $b$$=$$-0.265\pm0.034$ & & & \\
OLS(Y$\mid $X)     & 5.149$\pm$0.267 & 0.576$\pm$0.052 & 0.382$\pm$0.053 \\
OLS(X$\mid $Y)     & 7.158$\pm$0.417 & 0.967$\pm$0.081 & 0.641$\pm$0.086 \\
OLS bisector       & 6.060$\pm$0.270 & 0.753$\pm$0.052 & 0.499$\pm$0.063 \\
Orthogonal         & 5.719$\pm$0.330 & 0.687$\pm$0.064 & 0.455$\pm$0.064 \\
Reduced major axis & 6.024$\pm$0.276 & 0.747$\pm$0.054 & 0.495$\pm$0.063 \\
Mean OLS           & 6.153$\pm$0.286 & 0.772$\pm$0.056 & 0.511$\pm$0.065 \\
 & & & \\
$B$-band: $b$$=$$-0.200$ & & & \\
OLS(Y$\mid $X)     & 4.539$\pm$0.240 & 0.633$\pm$0.065 & 0.317$\pm$0.033 \\
OLS(X$\mid $Y)     & 6.321$\pm$0.347 & 1.112$\pm$0.093 & 0.556$\pm$0.047 \\
OLS bisector       & 5.326$\pm$0.225 & 0.844$\pm$0.060 & 0.422$\pm$0.030 \\
Orthogonal         & 5.134$\pm$0.298 & 0.793$\pm$0.080 & 0.397$\pm$0.040 \\
Reduced major axis & 5.305$\pm$0.234 & 0.839$\pm$0.063 & 0.420$\pm$0.032 \\
Mean OLS           & 5.430$\pm$0.238 & 0.872$\pm$0.064 & 0.436$\pm$0.032 \\
 & & & \\
$R$-band: $b$$=$$-0.278\pm0.040$ & & & \\
OLS(Y$\mid $X)     & 4.754$\pm$0.235 & 0.499$\pm$0.046 & 0.347$\pm$0.055 \\
OLS(X$\mid $Y)     & 6.434$\pm$0.360 & 0.825$\pm$0.071 & 0.573$\pm$0.089 \\
OLS bisector       & 5.532$\pm$0.244 & 0.650$\pm$0.048 & 0.452$\pm$0.067 \\
Orthogonal         & 5.127$\pm$0.284 & 0.571$\pm$0.056 & 0.397$\pm$0.065 \\
Reduced major axis & 5.489$\pm$0.248 & 0.642$\pm$0.049 & 0.446$\pm$0.067 \\
Mean OLS           & 5.594$\pm$0.256 & 0.662$\pm$0.050 & 0.460$\pm$0.069 \\
 & & & \\
$R$-band: $b$$=$$-0.200$ & & & \\
OLS(Y$\mid $X)     & 4.230$\pm$0.208 & 0.580$\pm$0.059 & 0.290$\pm$0.030 \\
OLS(X$\mid $Y)     & 5.727$\pm$0.294 & 1.003$\pm$0.084 & 0.502$\pm$0.042 \\
OLS bisector       & 4.902$\pm$0.201 & 0.770$\pm$0.058 & 0.385$\pm$0.029 \\
Orthogonal         & 4.663$\pm$0.258 & 0.702$\pm$0.074 & 0.351$\pm$0.037 \\
Reduced major axis & 4.876$\pm$0.208 & 0.762$\pm$0.060 & 0.381$\pm$0.030 \\
Mean OLS           & 4.978$\pm$0.211 & 0.791$\pm$0.060 & 0.396$\pm$0.030 \\
\end{tabular}

\medskip
Regression analysis between $\log v$ and 
$(\log h + b \mu_0)$ for the field galaxy data presented in Section 2
(excluding ESO-LV 1220040). 
The values $x$ and $y$ 
define the best-fitting plane such that $v \propto I_0^xh^y$ where 
$\mu_0 = -2.5\log I_0$. 
The TF relation is obtained when $b=-0.2$. 
\end{minipage}
\end{table*}

A significant fraction (1/2) of the LSB galaxies have rotational 
velocity estimates less than 100 km s$^{-1}$. 
In order to obtain an estimate as to how the LSB galaxies 
may be influencing the above results, the analysis was repeated 
using the HSB galaxy sample on their own, and then again using 
only the LSB galaxy sample.   Excluding the outliers 
identified in Figure~\ref{fig2}, 
the three-dimensional bisector method of regression for the HSB 
galaxy sample ($N=62, 61$) gave 
$v \propto I_0^{0.60\pm0.12}h^{0.82\pm0.12}$ ($B$-band) and
$v \propto I_0^{0.50\pm0.09}h^{0.66\pm0.09}$ ($R$-band).\footnote{From 
the second method of analysis, the `mixing' values were determined to be 
$b$$=$0.244$\pm$0.044 ($B$-band) and $b$$=$0.260$\pm$0.050 ($R$-band).}
One can therefore conclude that the LSB galaxies are not responsible for 
the observed, albeit slight, departures from the TF relation, nor are 
they biasing the analysis to produce values of $x$ as high as 0.5 and 
thereby causing the $M/L$ ratio to appear largely independent of 
surface brightness.  
%
%
The regression analysis of the LSB galaxy sample ($N=25, 27$) gave 
$v \propto I_0^{0.58\pm0.17}h^{0.92\pm0.15}$ ($B$-band) and
$v \propto I_0^{0.46\pm0.17}h^{0.79\pm0.18}$ ($R$-band), 
fully consistent with the result from the HSB galaxy sample. 
The reason for the slightly larger exponents when the HSB and the 
LSB galaxies are considered on their own, as opposed to taken 
together, can be easily understood by looking at the distribution 
of LSB and HSB galaxies in Figure~\ref{fig2}.  
This slight difference in slopes also illustrates how a selection 
bias against LSB galaxies would operate; removing the left half of 
the data in Figure~\ref{fig2} results in an increased slope in 
the best-fit to the remaining data (see, e.g., Lynden-Bell et al.\
1988, Appendix B). 

A remaining point of concern is that there may be relatively important 
distance errors for some of the more nearby galaxies due to 
streaming motions above the smooth Hubble flow.  
To address this concern, the regression analysis has been repeated 
excluding the 34 galaxies\footnote{Of the 5 galaxies identified 
previously as `outliers' in Figure~\ref{fig2}, only two of these 
are within this volume.} within 3,000 km s$^{-1}$.  
The results obtained from the regression
analysis of this data are fully consistent with the results from the 
larger galaxy sample, with 
$v \propto I_0^{0.53\pm0.06}h^{0.85\pm0.12}$ ($B$-band) and
$v \propto I_0^{0.45\pm0.05}h^{0.75\pm0.13}$ ($R$-band). 
Additionally, correction for Virgo-infall alone, using the 220-model of 
Kraan-Korteweg (1986), gave entirely consistent results with those
obtained using the more detailed velocity flow model of Burstein 
et al.\ (1989). 

A meaningful comparison of the best-fitting planes for the 
early-type and late-type ($>$Sc) spiral galaxies in the present data 
set is not possible because the uncertainty on the exponents is 
in general too great.  What we can, however, conclude at this point 
is that in the $B$-band, the near independency of $M/L_B$ on 
surface brightness holds when the HSB galaxies are taken on their 
own, when the HSB and LSB galaxies are taken together, and when 
one constrains the `mixing' value $b$ to -0.2 in order to mimic the 
TF relation.   Although, it is noted that this situation is expected 
to change in the redder passbands, where the slope of the TF relation 
is steeper (see section 5). 


\subsection{Does $M/L_B$ correlate with $\mu_{0,B}$?\label{MoL_Mu}}

\begin{figure}
\centerline{\psfig{figure=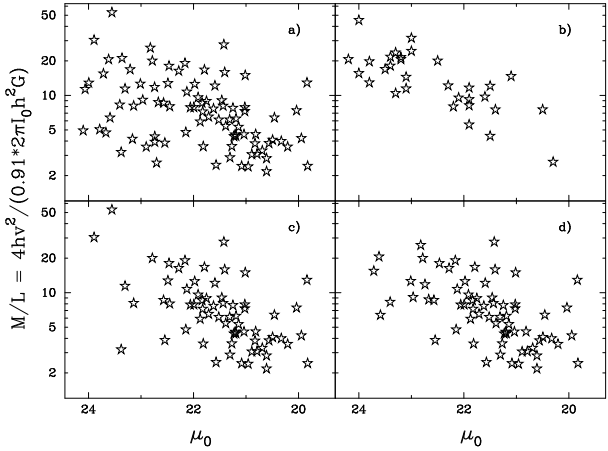,width=8.0cm,angle=-90}}
\caption{The kinematic $M/L$ ratio (in solar units) is shown against 
the central disk surface brightness ($B$-band data).
Panel a) contains the data described in Section 2, excluding the 
five outlying points identified in Figure~\ref{fig2}. 
Panel b) shows the smaller data sample from Table 1 of McGaugh \& 
de Blok (1998).  
Panel c) shows the data from panel a), excluding LSB galaxies with 
$\mu_0$ fainter than 22.65 $B$-mag arcsec$^{-2}$ and with $h$ smaller 
than 8.45 kpc (see Section~\ref{MoL_Mu} for details).
Panel d) shows the HSB galaxy sample from panel a) (excluding all three 
HSB galaxies with $\mu_0$ fainter than 23.0 $B$-mag arcsec$^{-2}$) 
together with the LSB galaxies in common with McGaugh \& de Blok (1998).
}
\label{figGaugh}
\end{figure}

Two points can be made regarding previous $B$-band studies which have 
claimed, and even plotted, a trend between mass-to-light ratio 
and central disk surface brightness.  
The first point is addressed by plotting the present data.
Figure~\ref{figGaugh}a shows the `kinematical' $M/L_B$ ratio 
(equation~\ref{MLat4})
versus the central $B$-band disk surface brightness $\mu_{0,B}$.  
One can see that although an upper 
envelope appears to exist, the current data present no strong correlation 
between $M/L_B$ and $\mu_{0,B}$.  Furthermore, the upper envelope is merely 
a reflection of the upper (bright) envelope in the $\mu_0$--$\log h$ 
diagram (Graham 2001b; Graham \& de Blok 2001) which is due to the 
absence of large scale-length galaxies with bright disks. 
Panel b) in Figure~\ref{figGaugh} displays the data from Table 1 
of McGaugh \& de Blok (1998).  
Their tabulated circular velocities, $B$-band scale-lengths and 
central disk surface brightnesses were used to compute the 
kinematical mass-to-light ratio within four scalelengths. 
This was done in exactly the same way as for panel a) 
and is in fact the method claimed to have been used in 
McGaugh \& de Blok (1998). 
Strangely, their tabulated $M/L$ ratios are different to the 
values derived from these basic parameters, and consequently 
their Figure 4b looks different to Figure~\ref{figGaugh}b 
shown here. 
Although the upper envelope is common between panels a) and b), 
the lower-left portions differ significantly.

The disk-dominated LSB galaxies from de Blok et al.\ (1995) have 
roughly the same scale-lengths as normal HSB spiral galaxies -- their 
median $B$-band scale-length is 4.3 kpc 
($H_{\rm 0}=75$ km s$^{-1}$ Mpc$^{-1}$; de Blok et al.\ 1995). 
The bulge-dominated LSB galaxies from Beijersbergen et al.\ (1999)
have a larger median $B$-band disk scale-length of 12.6 kpc 
(Beijersbergen et al.\ 1999).
Taking 8.45 kpc (=[12.6+4.3]/2) to represent some kind of division 
between the large and small (or rather, large and normal) sized LSB 
galaxies, panel c) shows 
the results after selecting against (i.e.\ excluding) those disks 
with central surface brightnesses more than one mag fainter than 
the canonical Freeman (1970) value of 21.65 $B$-mag and with scale-lengths 
less than 8.45 kpc.  These galaxies should roughly have, on average, 
the same scale-lengths as the HSB galaxy sample, but fainter surface 
brightnesses. 
If the $M/L$ ratio is independent of surface brightness, then 
we should find that we have removed those LSB galaxies with similar
$M/L$ ratios as the HSB galaxies - which indeed is what we do find.
Panel d) shows the current HSB galaxy sample (excluding the three 
disks with $\mu_{0}$ fainter than 23.0 $B$-mag arcsec$^{-2}$) together 
with those LSB galaxies in common with 
McGaugh \& de Blok (1998).  
From this restricted sample, a somewhat similar correlation between 
$M/L_B$ and $\mu_{0,B}$ to that present in the data of McGaugh 
\& de Blok (1998) appears. 
{\it The dependence of the $B$-band mass-to-light ratio on the central disk 
surface brightness can therefore arise from ones sample selection, 
even with a range of disk luminosities}

McGaugh \& de Blok (1998) excluded a number of LSB galaxies for a 
variety of different reasons.
All of the LSB galaxies in common with both studies
but excluded by McGaugh \& de Blok (1998) ($N=7$) reside very 
close to the mean relation defined by the other galaxies 
in Figure~\ref{fig2}.  That is, they do not appear to be outliers.  

The second point which should however be mentioned is that
in contradiction with most $B$-band TF studies, the small 
sample of McGaugh \& de Blok (1998) apparently has a TF slope of 
almost 4, i.e.\ $x=0.25$ (their Figure 1). 
In this case, as explained in Section 3, one does expect to find a strong 
correlation between $M/L$ and $\mu_0$, as these authors did for 
their galaxy sample. 



\section{Literature comparisons\label{lit_com}}

A somewhat similar three-dimensional analysis of 511 HSB Spiral 
(and Irregular) galaxies having types Sa through Im has been performed 
by Burstein et al.\ (1997) in the $B$-band.  
While they also used rotational velocity estimates from $W_{20}$ 
measurements, the nature of their photometric data is different 
to that used here.  Using the RC3 (de Vaucouleurs et al.\ 1991) tables 
of $B$-band circular effective aperture size and mean enclosed 
surface brightness, Burstein et al.\ (1997) used the axial-ratio-based 
statistical relationships in the RC3 to derive the mean effective surface 
brightness $<$$\mu$$>$$_e=-2.5\log$$<$$I_e$$>$ within the effective 
radius $r_e$.  They then corrected 
$<$$\mu$$>$$_e$ for Galactic extinction following Burstein \& Heiles (1982) 
and for internal extinction using the correction 1.5$\log(b/a)$. 
Recognizing errors of more than 0.1 dex in their 
photometric parameters, Burstein et al.\ (1997) chose not to apply 
the small $(1+z)^4$ correction or any $K$-correction term. 
Their investigation effectively used the total (bulge + disk) 
galaxy half-light radius and mean surface brightness within this radius. 
Depending on the contribution of light from the bulge, varying 
fractions of disk scale-length, and hence the disks themselves, 
were consequently sampled in their analysis. 
The present investigation differs as it has used one complete 
disk scale-length and the central disk surface brightness; 
an approach, it should be noted, which effectively regards the 
bulge matter as dark matter - the luminous flux from the bulge 
is ignored, while the presence of the bulge permeates into the 
analysis through the rotational velocity of the disk.   

If the relative contribution from the bulge light\footnote{$B/D$ 
ratios for much of the present galaxy sample have already
been presented in Graham (2001a,b) and Graham \& de Blok (2001).} 
is small enough that it can be neglected and the galaxy treated 
as simply an exponential disk, then the use of $r_e$ and 
$<$$\mu$$>$$_e$ (or $\mu_e$, the surface brightness at $r_e$) 
should yield similar results 
as the use of $h$ and $\mu_0$.  This is because $r_e=1.67h$ and 
$<$$\mu$$>$$_e=\mu_0 + 1.12$ ($\mu_e=\mu_0 + 1.82$) 
for an exponential disk.   
To investigate the influence of the bulge, and perform a somewhat 
more direct comparison with Burstein et al.\ (1997), the half-light
radius and the surface brightness at this 
radius have been derived for the present galaxy sample 
and the fundamental plane recomputed using these values.  The 
differences between the half-light radius of the disk (=1.67$h$) 
and the half-light radius of the galaxy ($r_e$), and the surface 
brightness of the disk at 1.67$h$ and of the galaxy at $r_e$ are 
shown in Figure~\ref{last}.  With the exception of a few 
galaxies\footnote{These galaxies are not the outliers identified in 
Figure~\ref{fig2}.} having very prominent bulges within their disk 
half-light radius, 
the disk and galaxy half-light radius usually varied by less than 
20 per cent, and the surface brightness terms by less than $\sim$0.4 mag 
arcsec$^{-2}$. The optimal fundamental plane constructed using 
$r_e$ and $\mu_e$ 
for the full sample (excluding the previously mentioned outliers) is
$v \propto I_e^{0.53\pm0.05}r_e^{0.79\pm0.09}$ 
($B$-band).   
For the HSB sample one finds 
$v \propto I_e^{0.61\pm0.11}r_e^{0.80\pm0.11}$, 
and for the LSB sample 
$v \propto I_e^{0.57\pm0.13}r_e^{0.87\pm0.17}$.

\begin{figure}
\centerline{\psfig{figure=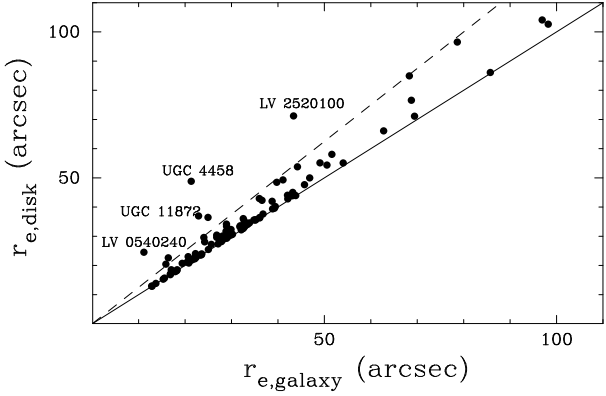,width=8.0cm,angle=-90}}
\caption{
a) The $B$-band half-light radii of the disks ($r_{\rm e,disk}=1.67h$) 
are plotted against the $B$-band half light radii of the galaxies 
($r_{\rm e,galaxy}$).  
The presence of the bulge accounting for the differences.  The 
dashed line denotes a 20 per cent difference from agreement (the solid line).
The marked outliers have very prominent bulges within the half-light
radius of the disk.  
b) The disk surface brightness at $r_{\rm e,disk}$ ($=\mu_0 + 1.82$) 
is plotted against the galaxy surface brightness at $r_{\rm e,galaxy}$.
}
\label{last}
\end{figure}

\begin{figure}
\centerline{\psfig{figure=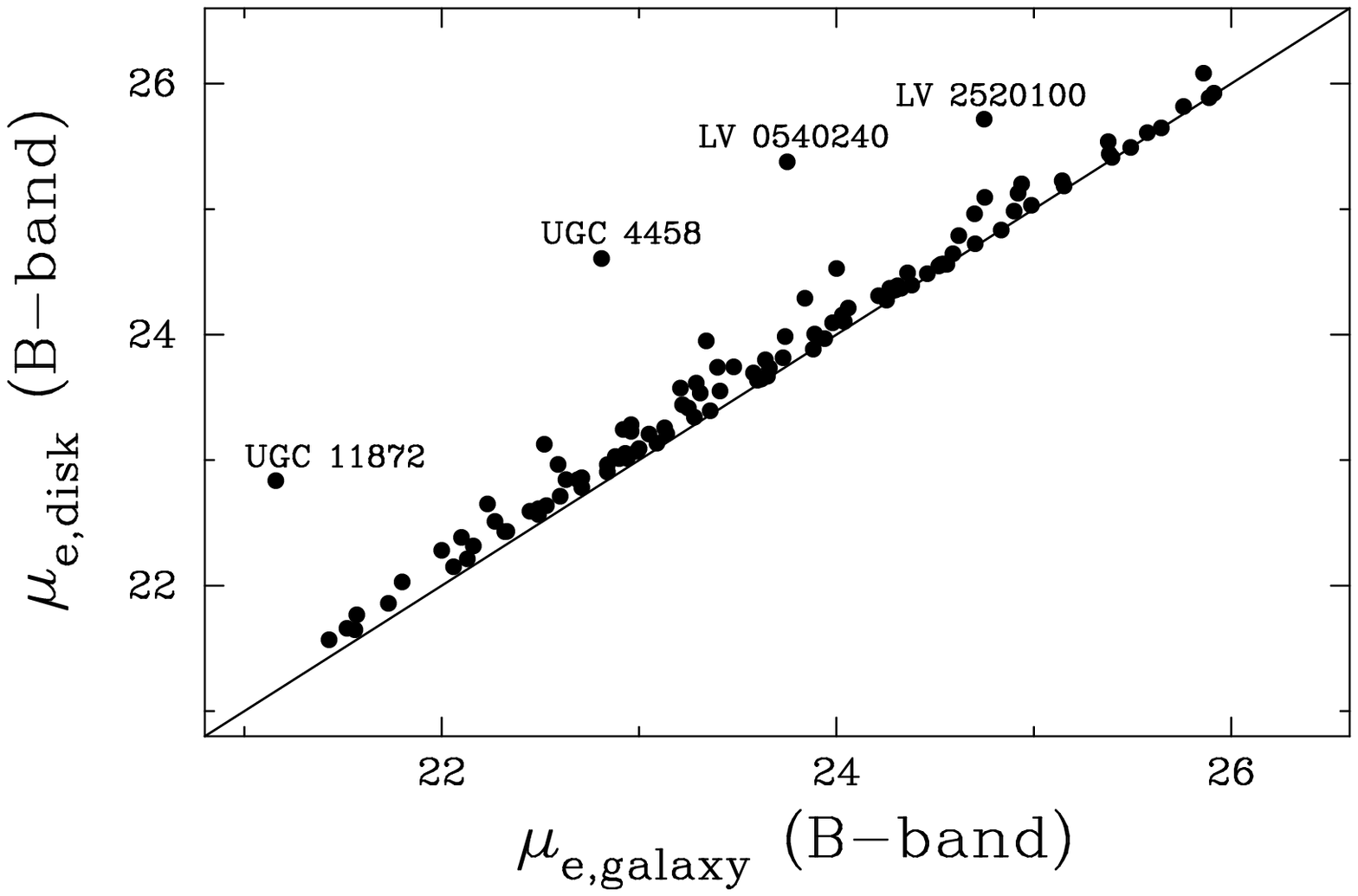,width=8.0cm,angle=-90}}
\contcaption{ }
\end{figure}

From the sample of 386 Sa-Sc Spiral galaxies analysed by 
Burstein et al.\ (1997), they found 
$v \propto <I>_e^{0.57\pm0.03}r_e^{0.85\pm0.05}$ ($B$-band). 
Applying the three-dimensional bisector regression 
analysis used in this paper to this very same data sample 
from Burstein et al.\ (1997) gives 
$v \propto <I>_e^{0.55\pm0.04}r_e^{0.78\pm0.05}$ ($B$-band).
From equation~\ref{eqMoL} this implies that 
$(M/L_B) \propto <I>_e^{0.10\pm0.08}r_e^{0.56\pm0.10}$,  
in good agreement with the present papers investigation. 

When Burstein et al.\ (1997) included all Spiral and Irregular
galaxy types in their analysis, they found 
$v \propto <I>_e^{0.39\pm0.02}r_e^{0.61\pm0.02}$ ($B$-band), 
a result with notably lower exponents.  Although, applying the 
three-dimensional bisector regression analysis gives 
$v \propto <I>_e^{0.41\pm0.01}r_e^{0.62\pm0.02}$, and 
excluding the Irregular type galaxies raises the exponents
slightly to give 
$v \propto <I>_e^{0.44\pm0.02}r_e^{0.66\pm0.02}$ ($B$-band). 
When the same internal extinction correction as used by 
Burstein et al.\ (1997) is applied to this papers field galaxy
sample, one obtains
$v \propto I_e^{0.48\pm0.04}r_e^{0.73\pm0.07}$ ($B$-band), 
consistent at the 1$\sigma$ level with the result from 
Burstein et al.\ (1997).

The result from Burstein et al.\ (1997), using all 511 disk galaxies and 
irregular galaxies, leads to (although not given in their paper) 
$L \propto v^{2.56\pm0.13}r_e^{0.44\pm0.07}$ ($B$-band).  
An even larger departure from the TF relation and dependence 
on a third parameter was found by Willick 
(1999) who derived $L \propto v^{2.4}r_e^{0.7}$ ($R$-band). 
Others have also previously found evidence for the residuals about 
the TF relation to correlate with a third parameter (Kodaira 1989; 
Feast 1994; Koda et al.\ 2000a; Han et al.\ 2001, Kannappan et al.\ 2002), 
and it may be that the TF relation reflects closely, 
but not exactly, the Fundamental-Plane-like distributions shown here.  
However, there are numerous papers (e.g.\ Strauss \& Willick 1995; 
Courteau 1997; and references therein) which, like this one, do not
detect deviations above the 1-2$\sigma$ level.  

It should also be kept in mind that different papers have used 
different methods of regression analysis to fit a plane or line to 
their data points.  The present papers interest lies 
in deriving $M/L$ ratio estimates which can be compared with 
theory.  The disk parameters have consequently been treated 
equally when constructing the optimal plane; an approach which
differs from the numerous direct and inverse TF relations 
constructed in the past for the determination of galaxy distances -- 
which is not the subject of the current paper.  A detailed 
comparison with every previous study is therefore not 
straight-forward if at all even possible.  However, a detailed 
analysis of one other well-respected data sample shall be presented. 

The reduced 
and tabulated photometric data of Tully et al.\ (1996) and the reduced 
kinematic data from Verheijen \& Sancisi (2001) is analysed. 
This consists of $B, R, I$ and $K'$ scale-lengths and central 
disk surface brightnesses, obtained from marking the disk by eye, and 
rotational velocities taken from where the rotation curve is flat 
($v_{\rm flat}$).  
The galaxy sample are all from the Ursa Major Cluster and consequently 
distance errors will not contribute quite as much to the scatter. 
For the sake of consistency, 
the surface brightness term has been uniformly corrected here in 
the same way as the previous data was dealt with (see Section 2). 
Because this new sample contains relatively edge-on Spiral galaxies, 
the $K'$-band inclination corrections to the central surface 
brightness terms are large, in some cases $>$2.0 mag arcsec$^{-2}$. 
For four galaxies which were catalogued to be completely edge-on, 
their inclinations were set to 85$^\circ$ when calculating this 
correction. 

\begin{figure}
\centerline{\psfig{figure=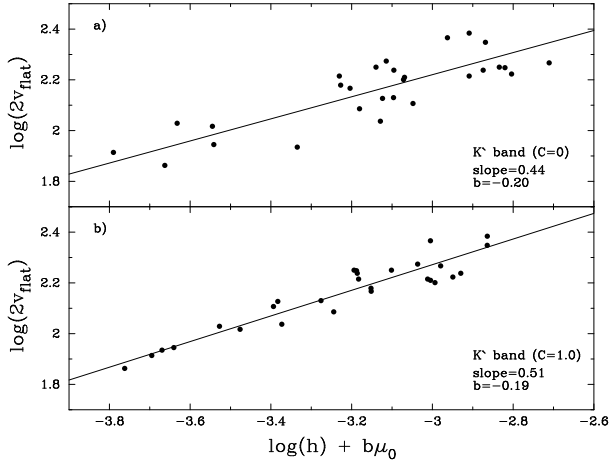,width=8.0cm,angle=-90}}
\caption{Orthogonal regression analysis to the $K'$-band data from 
Tully et al.\ (1996) and the velocity data from Verheijen \& 
Sancisi (2001).  
The value of $b$ was computed, as in Figure~\ref{fig2}, 
such that it resulted in the highest linear correlation 
coefficient between $\log 2v_{\rm flat}$ and $(\log h + b \mu_0)$. 
When $b=0.20$ these diagrams represent the 
TF relation, where $L\propto v^{2/{\rm slope}}$.  
The value of C dictates the apparent opacity of spiral galaxy 
disks for determinations of $\mu_0$, such that the inclination 
correction to the surface brightness term $\mu_0$ can be written 
as $2.5C\log(a/b)$, where $a/b$ is the observed outer disk axis ratio. 
Panel a) assumes the disks are opaque ($C=0$), while panel
b) assumes that they are fully transparent ($C=1$), 
resulting in considerably less scatter to the diagram. 
}
\label{figVer}
\end{figure}

The results of 
the three-dimensional bisector regression analysis are presented in 
Table~\ref{tab2}.  In the $B$-band, the analysis reveals that 
$\log(M/L_B) \propto (0.08\mu_0 + 0.46\log h)$, in good agreement 
with the previous $B$-band results and revealing that changes in 
scale-length are more important than changes in surface brightness.  
Figure~\ref{figVer} shows the results for the $K'$-band data 
with ($C=1.0$) and without ($C=0.0$) the inclination correction 
($2.5C\log[a/b]$, where $a/b$ is the observed disk axis-ratio) 
to the surface brightness term. 
From the scatter one can clearly see that at near-infrared 
wavelengths, spiral galaxy disks 
should be taken to be transparent for such tests (see also 
Graham 2001b). One can see that the TF relation is obtained 
(i.e.\ $y\sim 2x$), and the 
slope increases from the $B$-band to the $K'$-band such that 
$L_B\propto v^{2.5}$ and $L_{K'}\propto v^{4.0}$. 
The reduction in scatter with increasing wavelength has also 
been observed for many years (Aaronson, Huchra \& Mould 1979). 
Therefore, in the near-infrared, where one is affected little by 
obscuring dust and young stellar populations, the $M/L_{K'}$ 
ratio varies as $I_0^{-1/2}$, independent of $h$. 

\begin{table}
\caption{Regression Analysis: Cluster Galaxies}
\label{tab2}
\begin{tabular}{@{}cccc}
Band  &  $x$  &  $y$  &  $M/L$ \\[10pt]
$B$ &  0.396$\pm$0.059 & 0.732$\pm$0.118  & $I_0^{-0.208 \pm 0.118}h^{0.464 \pm 0.236}$ \\
$R$ &  0.334$\pm$0.035 & 0.642$\pm$0.098  & $I_0^{-0.332 \pm 0.070}h^{0.284 \pm 0.196}$ \\ 
$I$ &  0.303$\pm$0.029 & 0.601$\pm$0.092  & $I_0^{-0.394 \pm 0.058}h^{0.202 \pm 0.184}$ \\
$K'$ &  0.258$\pm$0.020 & 0.569$\pm$0.073  & $I_0^{-0.484 \pm 0.040}h^{0.138 \pm 0.146}$ \\
\end{tabular}

\medskip
Three-dimensional bisector regression analysis between 
$\log v$, $\mu_0$, and $\log h$ for Ursa Major Galaxy Cluster members.  
The exponents $x$ and $y$ 
define the best-fitting plane such that $v \propto I_0^xh^y$ where 
$\mu_0 = -2.5\log I_0$. 
The photometric data ($\mu_0, h$) comes from Tully et al.\ (1996) 
and the kinematic data ($v$) from Verheijen \& Sancisi (2001).
\end{table}

\begin{figure}
\centerline{\psfig{figure=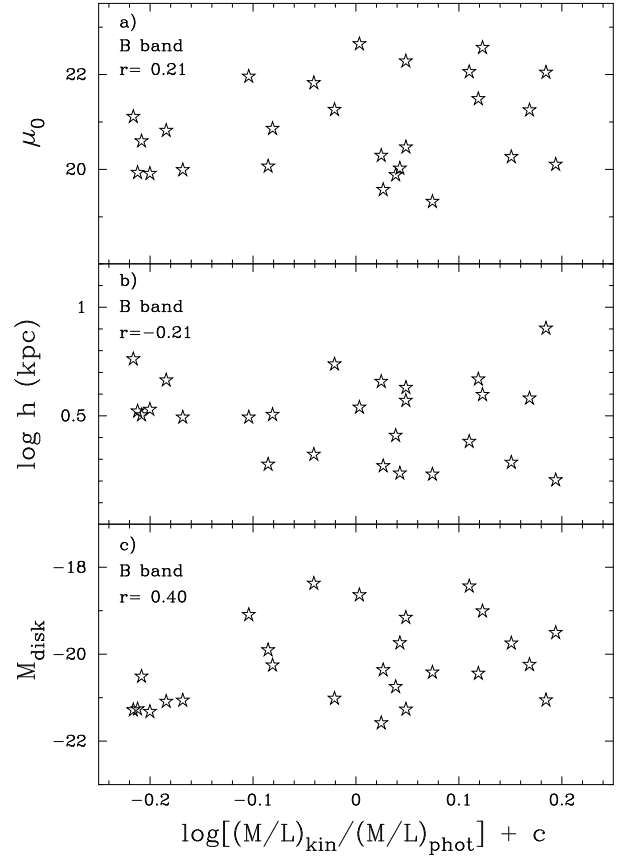,width=8.0cm,angle=00}}
\caption{
The logarithm of the difference between the mass-to-light ratio 
estimates obtained using equation~\ref{eqMoL} and equation~\ref{MLat4}
have been plotted against the central disk surface brightness (top 
panel), the logarithm of the disk scale-length (middle panel), and 
the absolute magnitude of the disk (bottom panel).
These plots show the $B$-band data from Tully et al.\ (1996) and 
the velocity data from Verheijen \& Sancisi (2001).
The values for $x$ and $y$ used
in equation~\ref{eqMoL} have been taken to be 0.396 and 0.732 (see 
Table~\ref{tab2}). 
Constants of proportionality have been ignored, and 
the value of $c$ in the $x$-axis is simply a constant. 
}
\label{figVB2}
\end{figure}

\begin{figure}
\centerline{\psfig{figure=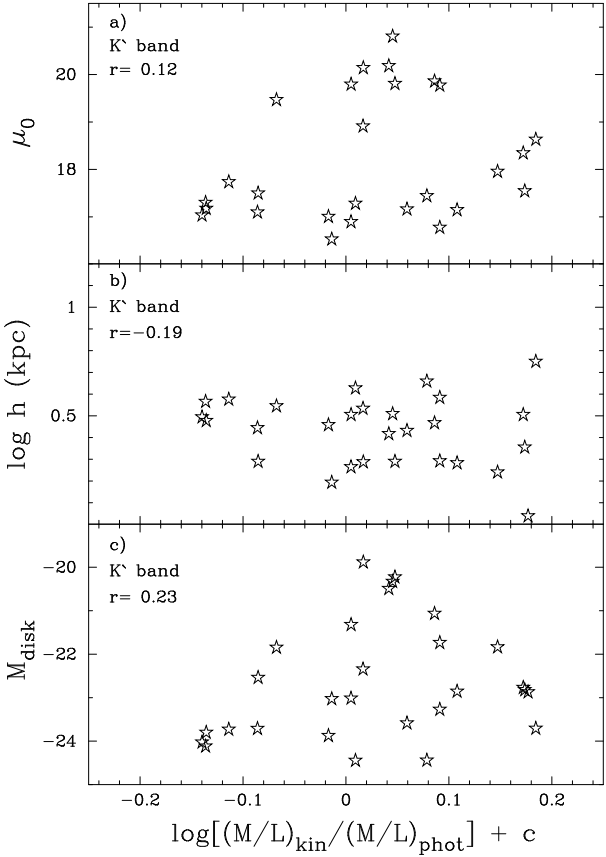,width=8.0cm,angle=00}}
\caption{
Same as Figure~\ref{figVB2} but using $K'$-band data. 
The values for $x$ and $y$ used
in equation~\ref{eqMoL} to derive $(M/L)_{\rm phot}$ have been taken 
to be 0.258 and 0.569.  
}
\label{figVK2}
\end{figure}

To again investigate possible differences between the mass-to-light
ratios obtained using equation~\ref{eqMoL} and equation~\ref{MLat4}, 
and hence possible signatures that the total-mass profile does not scale 
linearly with the luminous disk profile, we have repeated 
Figure~\ref{figML2} this time using the Ursa Major galaxy Cluster data.  
Figure~\ref{figVB2} presents the $B$-band analysis, which can be 
seen to display an identical behavior to that observed in 
Figure~\ref{figML2}.  This tends to suggest that the ratio of the 
mass profile scale-size to the $B$-band luminosity disk scale-length 
may not be constant, but scales weakly with the $B$-band disk 
luminosity (see Section 3).   Figure~\ref{figVK2} shows the
$K'$-band data where one can see that the weak $B$-band dependency 
on absolute magnitude ($r=0.4$, Figure~\ref{figVB2}, lower panel) is 
reduced even further.

Lastly, it should perhaps also be noted that there is another 
mechanism to change the slope of the optimal plane within a 
given passband.  
Tully et al.\ (1998) have presented evidence that the appropriate 
extinction correction for disk galaxies in Clusters 
may be a function of galaxy luminosity.  This was a subsequent 
paper to the Tully \& Verheijen (1997) paper which addressed the 
issue of extinction corrections and provided the somewhat more 
standard corrections which have been used in this paper.  
Tully et al.\ (1998) analysed the data from two galaxy Clusters 
(53 galaxies from Ursa Major and 34 galaxies from Pisces) to 
determine a TF slope of 3.1 in the B-band after applying their 
luminosity dependent extinction correction.  This is an increase 
of 0.2 from 2.9 found by Sprayberry et al.'s (1995) study of the 
Ursa Major Cluster.  Perhaps contrary to this, Willick et al.\ (1996) 
found no evidence for a luminosity-dependent extinction correction
when using field galaxies.  The interesting analysis by Tully et al.\ 
(1998) raises many questions.   For instance, the data for the Pisces 
Cluster galaxies on their own appear to suggest no luminosity 
dependent extinction correction.  Additionally, whether the results 
in Tully et al.\ (1998) can be extended to LSB galaxies is unclear; 
why an unevolved LSB galaxy with the same luminosity as a HSB galaxy 
should have the same amount of extinction due to dust is not obvious.  
Why the S0 and S0/a galaxies do not conform is a further mystery.  
The issue of corrections due to dust in field galaxies is not yet 
fully understood, or at least commonly accepted, and therefore it 
should be noted that the extinction corrections used here may need 
adjusting when this subject is on a firmer footing.

\section{Colours}

\begin{figure}
\centerline{\psfig{figure=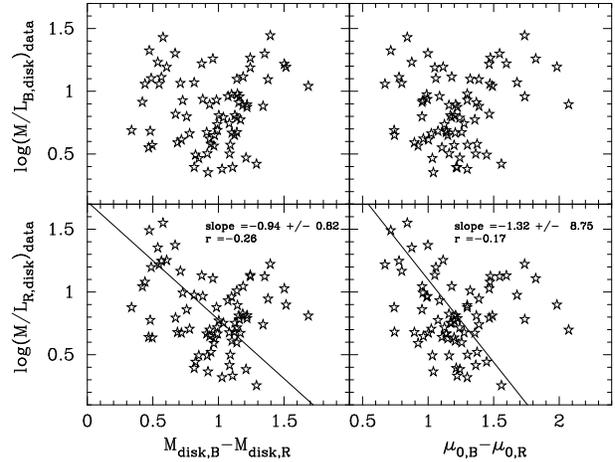,width=8.0cm,angle=-90}}
\caption{
The logarithm of the dynamical mass-to-light ratio obtained using equation 1 
(with $r=4h$) 
is plotted against two measures of the disk colour ($B-R$).  The left hand column
shows the colour difference between the integrated disk magnitude in $B$ and
$R$, while the right hand column shows the difference between the central
disk surface brightness in $B$ and $R$.  Only the disk luminosity (not the 
bulge and disk luminosity) has been 
used to derive the mass-to-light ratio, which is given here in solar units. 
An orthogonal regression analysis 
has been used to derive the slope (and error) to the correlation, and the 
Pearson correlation coefficient is also given for those plots displaying a 
possible correlation.
}
\label{figCo1}
\end{figure}

\begin{figure}
\centerline{\psfig{figure=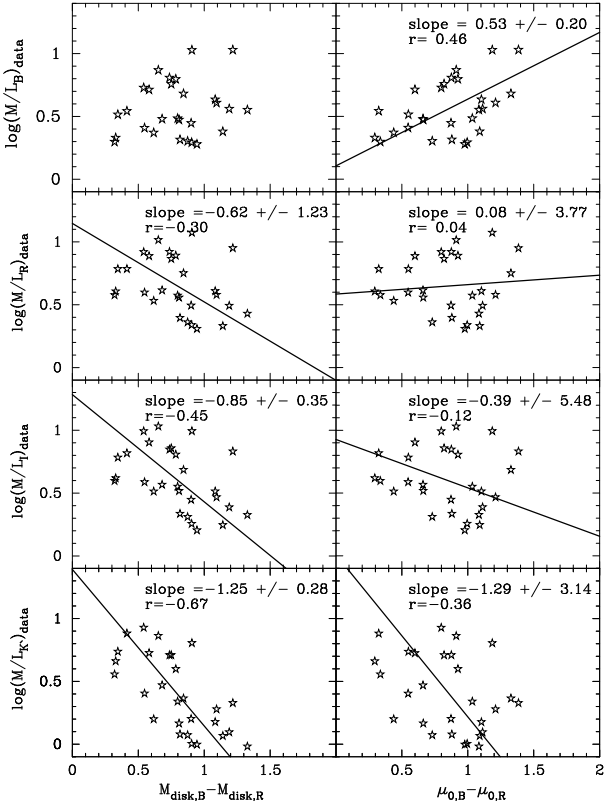,width=8.0cm,angle=00}}
\caption{
The same notation applies here as in Figure~\ref{figCo1}.  This time the 
Ursa Major galaxy Cluster data introduced in section 5 has been 
used. 
}
\label{figCo2}
\end{figure}

The ($B-R$) colour of each galaxy's disk has been estimated using two 
approaches.  Having previously modelled both the bulge and disk, 
we are better able to determine the ($B-R$) colour for the disk alone - 
rather than the integrated colour of the bulge and disk.  The first 
method has taken the difference between the corrected (see section 2) central 
disk surface brightness in $B$ and $R$, and the second approach has used 
the difference between the total disk magnitude determined 
in the $B$ and $R$ passbands.  Each galaxies dynamical mass-to-light 
ratio estimate (from equation~\ref{MLat4}) is plotted 
against its $B-R$ colour in Figure~\ref{figCo1} for the HSB and LSB field 
galaxies, and in Figure~\ref{figCo2} for the Ursa Major Cluster data 
from Tully et al.\ (1996) and Verheijen \& Sancisi (2001).   Only the disk 
light (that is, not the bulge light) has been used in this derivation of 
the dynamical mass-to-light ratio. 
Using the combined disk and bulge light to determine the 
mass-to-light ratio has very little effect on this result. 
It is immediately obvious that, within the noise of the present 
field galaxy data, 
no obvious correlation exists between the $B$- and $R$-band dynamical 
mass-to-light ratio and $B-R$ colour. 
An orthogonal regression analysis has been performed for those 
plots showing a possible correlation, and the Pearson correlation 
coefficient determined (results shown within the figures).  
In general, the correlations are weak.  


Using a suite of simplified galaxy disk evolution models, Bell \& 
de Jong (2001) have presented the case that the {\it stellar} 
mass-to-light ratio of the disk ($M_{*,{\rm disk}}/L_{\rm disk}$) 
should be correlated with colour.  Their strongest 
correlation was obtained with the colour ($B-R$) such that the logarithm 
of a disks $B$-band stellar mass-to-light ratio varies as 1.25($B-R$).   
While the present paper has dealt with the dynamical, or total, 
mass-to-light ratio ($M_{\rm dyn}/L_{\rm disk}$) within a fixed number 
of disk scale-lengths (i.e.\ 4), one can 
combine the present observational results with the theoretical 
predictions of Bell \& de Jong (2001) to determine how the 
stellar-to-dynamical\footnote{By stellar-to-dynamical mass ratio, 
it is meant the ratio of the stellar disk mass (excluding any possible
bulge mass) to the total dynamical mass within 4 scale-lengths.} 
mass ratio must vary with ($B-R$) colour.  

The only strong correlation detected between the dynamical 
mass-to-light ratio and ($B-R$) colour is in the $K'$-band 
data of the Cluster Spiral galaxies (Figure~\ref{figCo2}), where 
$\log(M_{\rm dyn}/L_{K'{\rm disk}}) \propto -1.25\pm0.28(B-R)$  
(Figure~\ref{figCo2}).   Combining this result with the stellar 
mass-to-light ratio prediction of Bell \& de Jong (2001)
($\log[M_{*,{\rm disk}}/L_{K,{\rm disk}}] \propto 0.45[B-R]$),
and neglecting differences between $K$ and $K'$, would imply that 
the logarithm of the stellar-to-dynamical mass ratio must increase 
as disks become redder, varying as 1.70$\pm0.28$($B-R$).  
If the theoretical models are correct, {\it this implies that 
galaxy modellers should not assume a constant stellar-to-total 
mass ratio (at least within 4$h$) for any given passband.}
This conclusion seems to be at least qualitatively correct; 
LSB galaxies are largely unevolved blue galaxies with higher 
relative gas-to-total mass fractions than HSB galaxies 
(Bothun, Impey \& McGaugh 1997).

\section{Conclusions}


The best-fitting two-dimensional plane to the 3-dimensional space of 
galaxy disk parameters (rotational velocity $v_{\rm rot}$, disk 
scale-length $h$, and central disk surface brightness 
$\mu_0=-2.5\log I_0$) has been derived for a sample of nearly 
100 field galaxies.  In the $B$-band 
$v_{\rm rot}\propto I_0^{0.50\pm0.05}h^{0.77\pm0.07}$, and in the 
$R$-band $v_{\rm rot}\propto I_0^{0.43\pm0.04}h^{0.69\pm0.07}$. 
This implies that in the $B$-band the dynamical $M/L$ ratio is, 
contrary to popular claims, nearly independent of the disk surface 
brightness level; the disk scale-length is what is important, with 
$\log(M/L_B) \propto 0.00(\pm0.04)\mu_0 + 0.54(\pm0.14)\log h$. 
This general result holds when using HSB field galaxies, 
when using HSB and LSB field galaxies together, and when 
derived from the TF relation instead of the best-fitting 
two-dimensional plane. 
We should therefore not talk simply about LSB and HSB 
galaxies, but also give mention to their scale-lengths. 

The $M/L_B$ ratio is shown to be largely independent of a 
disks ($B-R$) colour, whereas $M/L_K'$ varies as -1.25$\pm$0.28($B-R$). 
Combining this with recent theoretical modelling implies that 
the logarithm of the stellar-to-dynamical 
mass ratio is not a constant value, but increases as disks become 
redder, varying as 1.70$\pm0.28$($B-R$).  Consequently, galaxy 
modellers should not treat this as a constant quantity.  

In the $R$-band 
$\log(M/L_R) \propto 0.06(\pm0.03)\mu_0 + 0.38(\pm0.14)\log h$. 
Extensions to redder passbands, albeit using a sample of Cluster 
spirals, reveals that the $M/L$ ratio becomes increasingly less 
dependent on the disk scale-length at longer wavelengths, reaching 
the situation where $M/L \propto 1/\sqrt{I_0}$ at $K'$. 

Marginal evidence is presented to suggest that the ratio of the 
mass-profile scale-size ($a$) to the luminous disk scale-length ($h$) 
may vary with disk luminosity in the $B$-band, such that $a/h$ 
decreases with increasing disk luminosity.  A method to test this 
tentative possibility is described.

\section{acknowledgements}
I wish to thank Erwin de Blok for kindly providing me with electronic 
tables of the kinematical data used in this study. 
I am also grateful to David Burstein and Ren\'{e}e Kraan-Korteweg for 
providing me with their velocity flow codes, and to Bahram Mobasher 
and Ken Freeman for their careful reading and helpful comments on 
this manuscript. 

This research has made use of the LEDA database 
(http://leda.univ-lyon1.fr).  
This research has made use of the NASA/IPAC Extragalactic Database (NED)
which is operated by the Jet Propulsion Laboratory, California Institute
of Technology, under contract with the National Aeronautics and Space
Administration.
The Parkes telescope is part of the Australia Telescope which is funded
by the Commonwealth of Australia for operation as a National Facility
managed by CSIRO.

\label{lastpage}


\begin{thebibliography}{99999}
\bibitem{ahm79}Aaronson M., Huchra J., Mould J., 1979, ApJ, 229, 1
\bibitem{Aet01}Andersen D.R., Bershady M.A., Sparke L.S., Gallagher J.S., Wilcots E.M., 2001, ApJ, 551, L131
\bibitem{Bet01}Barnes D.G., et al., 2001, MNRAS, 322, 486
\bibitem{Bdv99}Beijersbergen M., de Blok W.J.G., van der Hulst J.M., 1999, A\&A, 351, 903
\bibitem{BaJ01}Bell E.F., de Jong R.S., 2001, ApJ, 550, 212 
\bibitem{BIM97}Bothun G.D., Impey C.D., McGaugh S.S., 1997, PASP, 109, 745 
\bibitem{BIS01}Burkholder V., Impey C., Sprayberry D., 2001, AJ, 122, 2318
\bibitem{BBF97}Burstein D., Bender R., Faber S.M., Nolthenius R., 1997, AJ, 114, 1365
\bibitem{Bur89}Burstein D., Davies R.L., Dressler A., Faber S.M., Lynden-Bell D., 1989, in Large scale structure and motions in the universe, Mezetti, M., ed., Dordrecht, Kluwer Academic Publishers, p.179 
\bibitem{CCD93}Caon N., Capaccioli M., D'Onofrio M., 1993, MNRAS, 265, 1013
\bibitem{DSS97}Dalcanton J.J., Spergel D.N., Summers F.J., 1997, ApJ, 482, 659
\bibitem{daM97}de Blok W.J.G., McGaugh S.S., 1997, MNRAS, 290, 533
\bibitem{dMR01}de Blok W.J.G., McGaugh S.S., Rubin V.C., 2001, ApJ, in press
\bibitem{dMv96}de Blok W.J.G., McGaugh S.S., van der Hulst J.M., 1996, MNRAS, 283, 18 
\bibitem{dvB95}de Blok W.J.G., van der Hulst J.M., Bothun G.D., 1995, MNRAS, 274, 235
\bibitem{Cou97}Courteau S., 1997, AJ, 114, 2402 
\bibitem{deJ96}de Jong R.S., 1996, A\&AS, 118, 557
\bibitem{dav94}de Jong R.S., van der Kruit P.C., 1994, A\&AS, 106, 451
\bibitem{deV91}de Vaucouleurs G., de Vaucouleurs A., Corwin H.G., Buta R.J., Paturel G., Fouque P., 1991, Third Reference Catalog of Bright Galaxies (Springer-Verlag, New York)(RC3)
\bibitem{DaD87}Djorgovski S., Davis M., 1987, ApJ, 313, 59
\bibitem{Det87}Dressler A., Lynden-Bell D., Burstein D., Davies R.L., Faber S.M., Terlevich R.J., Wegner G., 1987, ApJ, 313, 42
\bibitem{FaJ76}Faber S.M., Jackson R.E., 1976, ApJ, 204, 668
\bibitem{Fet87}Faber S.M., Dressler A., Davies R.L., Burstein D., Lynden-Bell D., Terlevich R.J., Wegner G., 1987, in Faber S.M., ed., Nearly Normal Galaxies, Springer, New York, p.175
\bibitem{Fea94}Feast M.W., 1994, MNRAS, 266, 255
\bibitem{FaB92}Feigelson E.D., Babu G.J., 1992, ApJ, 397, 55
\bibitem{FdZ92}Franx M., de Zeeuw T., 1992, ApJ, 392, L47
\bibitem{Fre70}Freeman K.C., 1970, ApJ, 160, 811
\bibitem{Gr00b}Graham A.W., 2001a, AJ, 121, 820 
\bibitem{Gr00a}Graham A.W., 2001b, MNRAS, in press
\bibitem{GaC97}Graham A.W., Colless M.M., 1997, MNRAS, 287, 221
\bibitem{GdB01}Graham A.W., de Blok W.J.G., 2001, ApJ, in press
\bibitem{Han01}Han J.L., Deng Z.G., Zou Z.L., Wu X.B., Jing Y.P., 2001, PASJ, 53, 853
\bibitem{Het98}Haynes M.P., van Zee L., Hogg D.E., Roberts M.S., Maddalena R.J., 1998, AJ, 115, 62
\bibitem{Het00}Huchtmeier W.K., Karachentsev I.D., Karachentseva V.E., Ehle M., 2000, A\&AS, 141, 469
\bibitem{Iet90}Isobe T., Feigelson E.D., Akritas M.G., Babu G.J., 1990, ApJ, 364, 104
\bibitem{Kod0a}Koda J., Sofue Y., Wada K., 2000a, ApJL, 531, L17
\bibitem{Kod0b}Koda J., Sofue Y., Wada K., 2000b, ApJ, 532, 214
\bibitem{Kod89}Kodaira K., 1989, ApJ, 342, 122 
\bibitem{RKK86}Kraan-Korteweg R.C., 1986, A\&AS, 66, 255
\bibitem{Ket98}Kravtsov A.V., Klypin A., Bullock J.S., Primack J.R., 1998, ApJ, 502, 48
\bibitem{LaV89}Lauberts A., Valentijn E.A., 1989, The Surface Photometry Catalogue of the ESO-Uppsala Galaxies (ESO-LV). European Southern Observatory
\bibitem{LYJ86}Lilje P.B., Yahil A., Jones B.J.T., 1986, ApJ, 307, 91
\bibitem{Lyn88}Lynden-Bell D., Faber S.M., Burstein D., Davies R.L., Dressler A., Terlevich R.J., Wegner G., 1988, ApJ, 326, 19
\bibitem{MaG02}Matthews L.D., Gallagher J.S., ApJS, in press (astro-ph/0203188) 
\bibitem{MaM00}Mo H.J., Mao S., 2000, MNRAS, 318, 163
\bibitem{MMW98}Mo H.J., Mao S., White S.D.M., 1998, MNRAS, 295, 319
\bibitem{Met99}McGaugh S.S., Schombert J.M., Bothun G.D., de Blok W.J.G., 1999, ApJ, 533, L99
\bibitem{Mad99}McGaugh S.S., de Blok W.J.G., 1998, ApJ, 499, 41
\bibitem{NFW96}Navarro J.F., Frenk C.S., White S.D.M., 1996, ApJ, 462, 563
\bibitem{Pog97}Poggianti B.M., 1997, A\&AS, 122, 399
\bibitem{RaS94}Richter O.-G., Sancisi R., 1994, A\&A, 290, 9
\bibitem{Ret85}Rubin V.C., Burstein D., Ford W.K.Jr., Thonnard N., 1985, ApJ, 289, 81
\bibitem{SFD98}Schlegel D.J., Finkbeiner D.P., Davis M., 1998, ApJ, 500, 525
\bibitem{SaB88}Schombert J.M., Bothun G.D., 1988, AJ, 95, 1389
\bibitem{Set92}Schombert J.M., Bothun G.D., Schneider S.E., McGaugh S.S., 1992, AJ, 103, 1107
\bibitem{Ser68}S\`ersic J.-L., 1968, Atlas de Galaxias Australes (Cordoba: Observatorio Astronomico)
\bibitem{Set02}Kannappan S.J, Fabricant D.G., Franx M., 2002, AJ, in press (astro-ph/0202111)
\bibitem{SaW96}Strauss M.A., Willick J.A., 1995, Phys.\ Rep., 261, 271
\bibitem{SMT00}Swaters R.A., Madore B.F., Trewhella M., 2000, ApJ, 531, L107
\bibitem{Tet00}Tonry J.L., Blakeslee J.P., Ajhar E.A., Dressler A., 2000, 530, 625
\bibitem{TLR94}Triay R., Lachieze-Rey M., Rauzy S., 1994, A\&A, 289, 19 
\bibitem{TaF77}Tully R.B., Fisher J.R., 1977, A\&A, 54, 661
\bibitem{TaF85}Tully R.B., Fouqu\'e, P., 1985, ApJS, 58, 67
\bibitem{Tet98}Tully R.B., Pierce M.J., Jia-Sheng H., Saunders W., Verheijen M.A.W., Witchalls P.L., 1998, AJ, 115, 2264
\bibitem{Tet96}Tully R.B., Verheijen M.A.W., Pierce M.J., Jia-Sheng H., Wainscoat R.J., 1996, AJ, 112, 2471 
\bibitem{TaV97}Tully R.B., Verheijen M.A.W., 1997, ApJ, 484, 145
\bibitem{vdB00}van den Bosch, F.C., 2000, ApJ, 530, 177
\bibitem{vet00}van den Bosch, F.C., Robertson B.E., Dalcanton J.J., de Blok W.J.G., 2000, AJ, 119, 1579
\bibitem{vdH93}van der Hulst J.M., Skillman E.D., Smith T.R., Bothun G.D., McGaugh S., de Blok W.J.G., 1993, AJ, 106, 548
\bibitem{Vet01}Verheijen M.A.W., Sancisi R., 2001, A\&A, in press (astro-ph 0101404)
\bibitem{Wil99}Willick J.A., 1999, ApJ, 516, 47 
\bibitem{Wet96}Willick J.A., Courteau S., Faber S.M., Burstein D., Dekel A., Kolatt T., 1996, ApJ, 457, 460
\bibitem{Wet97}Willick J.A., Courteau S., Faber S.M., Burstein D., Dekel A., Strauss M.A., 1997, ApJS, 109, 333
\bibitem{Zet95}Zwaan M.A., van der Hulst J.M., de Blok W.J.G., McGaugh S.S., 1995, MNRAS, 273, L35
\end{thebibliography}
\end{document}